\documentclass[12pt]{iopart}
\pdfoutput=1
\usepackage{iopams}
\usepackage{graphicx}
\usepackage{subfigure}
\usepackage{xcolor}
\usepackage{soul}

\expandafter\let\csname equation*\endcsname\relax
\expandafter\let\csname endequation*\endcsname\relax
\usepackage{amsmath}

\usepackage{wasysym}
\usepackage{amssymb}
\usepackage{color}
\usepackage{soul}
\usepackage{ulem}

\expandafter\let\csname equation*\endcsname\relax 
 \expandafter\let\csname endequation*\endcsname\relax 
\begin{document}

\title{Networks with preferred degree: A mini-review and some new results}
\author{Kevin E. Bassler$^{1,2,3}$, Deepak Dhar$^{4}$, and R. K. P. Zia $%
^{3,5,6} $}

\address{$^1$ Department of Physics, Department of
Physics, University of Houston, Houston, TX 77204, USA}%
\address{$^2$Texas
Center for Superconductivity, University of Houston, Houston, TX 77204, USA}%
\address{$^3$ Max-Planck-Institut f\"{u}r Physik komplexer Systeme, N\"{o}thnitzer
Str. 38, Dresden D-01187, Germany}%
\address{$^4$ Department of Theoretical Physics 
Tata Institute of Fundamental Research, Homi Bhabha Road, Mumbai 400005, India}%
\address{$^5$Department of Physics and Astronomy, Iowa State University,
Ames, IA 50011, USA}%
\address{$^6$ Department of Physics,
Virginia Polytechnic Institute and State University, Blacksburg, VA 24061, USA}

\begin{abstract}
Since their inception about a decade ago, dynamic networks which adapt to
the state of the nodes have attracted much attention. One simple case of
such an adaptive dynamics is a model of social networks in which individuals
are typically comfortable with a certain number of contacts, i.e., preferred
degrees. This paper is partly a review of earlier work of single homogeneous
systems and ones with two interacting networks, and partly a presentation of
some new results. In general, the dynamics does not obey detailed balance
and the stationary distributions are not known analytically. A particular
limit of the latter is a system of extreme introverts and extroverts - the $%
XIE$ model. Remarkably, in this case, the detailed balance condition is
satisfied, the exact distribution and an effective Hamiltonian can be found
explicitly. Further, the model exhibits a phase transition in which the
total number of links in the system - a macroscopically interesting
quantity, displays an extreme Thouless effect. We show that in the limit of
large populations and away from the transition, the model reduces to one
with non-interacting agents of the majority subgroup. We determine the
nature of fluctuations near the transition. We also introduce variants of
the model where the agents show preferential attachment or detachment. There
are significant changes to the degree distributions in the steady state,
some of which can be understood by theoretical arguments and some remain to
be explored. Many intriguing questions are posed, providing some food for
thought and avenues for future research.
\end{abstract}

\ead{bassler@uh.edu, ddhar@theory.tifr.res.in, rkpzia@vt.edu}


\maketitle

\section{Introduction}

Social networks are common in nature and show a fascinating variety of
complex, collective behaviors. They occur in systems that range from the
microscopic to the gigantic, e.g., quorum sensing bacteria \cite%
{MillerBassler01}, colonies of ants \cite{Ants}, starling murmurations \cite%
{Swarms}, and whale pods \cite{Whales}. Often, the `bonds' between
individuals are not directly observable, and so properties of the `network'
must be inferred from the \textit{correlated} behavior of individuals. Thus,
understanding the properties of social networks poses serious challenges.
Among all such networks, human social networks are arguably the most
complex. To describe them adequately and appropriately is already extremely
difficult, let alone understanding them and predicting their evolution

In this effort, the study of simple models that show some collective effects
can be quite instructive. This is also a prerequisite for tackling more
complex and realistic models. In this spirit, this article is devoted to a
simple model of a \textit{dynamic} social network which incorporates the
notion that a person tends to add (cut) contacts when they have fewer (more)
than they would like to have. For simplicity, we will model this tendency by
assigning a number to each node, $\kappa $, the \textit{preferred degree}. 
Of course, $\kappa $ can vary from one individual to another and also, from
time to time. The value of $\kappa $ may be determined internally or imposed
externally. Examples of the former include introverts who prefer few friends 
\textit{vs.} extroverts who prefer many, while medical quarantine is a good
example of the latter. Incorporating some randomness in the actions of
individuals, the network would undergo a stochastic evolution. What are the
statistical properties of such a network (e.g., degree distribution)? Are
they mundane? or surprising? Are some properties more robust, independent of
details of evolution rules, similar to the universality seen in equilibrium
phase transitions?

As a first step, we consider the behavior of a homogeneous population, in
which everyone has the same, time-independent $\kappa $. Obviously, the
general population is more diverse, characterized by a distribution of $%
\kappa $s, and with some variability in time. Thus, our next step is to
introduce diversity in the simplest way: each individual, or agent, has one
of two $\kappa $s. For such a system, we may refer to the two subgroups as
`introverts' ($I$) and `extroverts' ($E$) with $\kappa _{I}<\kappa _{E}$.
Even in this simple model, there are very many ways to couple the two
`communities' together. What can we learn from the different ways one
subculture interacts with another? In the rest of this article, we will
describe a number of reasonably realistic scenarios, though we will study in
detail only a few. Not surprisingly, some aspects of the collective
behavior, first seen in simulations, can be understood with hindsight, while
others may appear to be counter-intuitive and are quite difficult to
analyze. Once a few baselines are established and understood,
generalizations can be incorporated and more complex models can be
investigated.

Though we will focus only on the properties of a fluctuating network in
which the nodes have no degrees of freedom, realistic and interesting social
structures can consist of nodes with their own variables, e.g., health, wealth,
and opinion. Frequently, these degrees of freedom feed back to the dynamics
of the network. For example, a sick individual is more likely to stay home
and so, have fewer contacts than when he/she recovers. We can model such
situations by letting $\kappa $ be \textit{dependent} on the state of the
person. Alternatively, a healthy individual may prefer to stay home when
he/she learns of an ongoing epidemic in the community. In this case, we let 
$\kappa $ depend on the state of the whole population 
\cite{JoladLiuSchmittmannZia12}. 
Such, so-called, adaptive networks, in which the
nodes and the links `co-evolve,' describe many important biological and
social systems \cite{1,2,3,4,5,6}. Here, we will restrict ourselves to
systems with `static nodes and dynamic links.' In this vein, we should
mention that, in theoretical condensed matter physics, conventional studies
deal with `dynamic nodes and static links.' For example, in the textbook
Ising model, the network that specifies the interactions is fixed
(typically, a regular lattice in some dimension). Even in theories of
strongly correlated electrons, the interactions between them do not
fluctuate randomly as a function of the state of the electrons.

The outline of the paper is as follows: Section 2 is devoted to a brief
review, highlighting the more surprising aspects of our discoveries in \cite%
{CSP24,CSP25,LiuSchmittmannZia12,LiuJoladSchZia13,LiuSchZia14,BasslerLSZ15,WJLthesis}%
. This review, including the material in Appendix A, is designed
pedagogically for students and non-experts, exposing the basic formulation
of dynamic networks, as well as certain principal characteristics of the
non-equilibrium stationary states they settle into. For completeness, we
first provide detailed descriptions of the dynamics of networks with
preferred degree and the master equation associated with the stochastic
process. These involve a baseline model (a homogeneous population), as well
as somewhat more realistic systems with two subgroups, interacting via a
variety of ways. Discovered through simulations, much of the statistical
properties of various degree distributions (a standard observable associated
with networks) can be understood through simple arguments and mean-field
approaches. In this effort, the study of a particular limiting case of the
above has been particularly rewarding. Consisting of e\underline{\textit{x}}%
treme \underline{\textit{i}}ntroverts and \underline{\textit{e}}xtroverts
who prefer contact with no one and everyone, respectively, it has been
called the $XIE$ model. For this special, analytically tractable limit, we
found the exact stationary distribution, as well as good approximations for
most degree distributions. 
More significantly, this system exhibits the characteristics of both a first
and second order transition. Known as the Thouless effect \cite%
{Thouless,Aize88,Luij01}, which has been observed only in equilibrium
systems \cite{Blossey95,Poland66,Fisher66,Kafri00,Gross85,Schwarz06,Toni06},
the order parameter suffers a discontinuity across the transition and
displays anomalously large fluctuations. Indeed, our system shown an extreme
form of this effect \cite{BarMukamel14,BarMukamel14a}, as our equivalent of
the magnetisation jumps from $-1$ to $1$, while wandering over the entire
interval $[-1,1]$ at the transition. 
In Section 3, we show that, in the limit of large number of agents ($N$) and
away from phase transitions, the $XIE$ model becomes exactly soluble, as the
agents in the majority become effectively independent, with residual
interactions vanishing as $N^{-1/2}$. We also provide a scaling theory for
the fluctuations in a certain neighborhood, or scaling window, of the
critical point. In section 4, we introduce two variants of the $XIE$ model,
where the agents are more selective with which links to add or cut. Novel
behavior of the steady states are discovered through simulations, some of
which can be understood theoretically. We end with a summary and outlook in
Section 5.

\section{Dynamic networks with preferred degrees: a brief review}

In all the models described in this article, we consider a population of $N$
individuals (labelled by $i=1,...,N$), each associated with a preferred
degree $\kappa \left( i\right) $. We begin with systems in which the agents
are not endowed with any degree of freedom, so that only the connections
between them are dynamic - i.e., static nodes and dynamic links. The
network, specified by an adjacency matrix $\mathbb{A}$ (with elements $%
A_{ij}=A_{ji}=1,0$ if nodes $i$ and $j$ are connected or not, respectively,
and $A_{ii}\equiv 0$ to exclude self loops), evolves as follows: At each
time step, a random agent is chosen and its degree, $k_{i}=\Sigma _{j}A_{ij}$%
, is noted. If $k_{i} \geq \kappa \left( i\right) $, it chooses randomly one
of its links to cut. Otherwise, it adds a link to a randomly chosen partner
not already connected to it. \footnote{%
Note that these rules prevent the system from having an absorbing state,
which would be the case if we let an agent with exactly $\kappa $ links do
nothing.} This rule, though not so realistic, models the individual's
attempt to restore its degree towards the preferred $\kappa $. Of course, we
can soften this `rigid' rule, by specifying smoother functions of $k$ for
the probabilities $w_{\pm }\left( k;\kappa \right) $, with which the agent
will add/cut a link, given that it has $k$ and prefers $\kappa $. However,
for most of the models we study in detail, we will use the rigid rule, for
simplicity. Thus, the total number of links in the system changes by unity
at each time step (for $0<\kappa <N-1$). The stochastic process of the
entire network is described in terms of $\mathcal{P}(\mathbb{A},t~|\mathbb{A}%
_{0},0)$, which is the probability of finding configuration $\mathbb{A}$ at
time $t$, given an initial configuration $\mathbb{A}_{0}$. As our main
interest is in stationary states, we can ignore the initial state and
simplify our notation to $\mathcal{P}(\mathbb{A},t)$. The rules governing
its evolution are embodied in a discrete master equation: 
\begin{equation}
\mathcal{P}(\mathbb{A}^{\prime },t+1)=\sum_{\mathbb{A}}\mathcal{R}(\mathbb{%
A\rightarrow A}^{\prime })\mathcal{P}(\mathbb{A},t)  \label{MEq}
\end{equation}%
where $\mathcal{R}(\mathbb{A\rightarrow A}^{\prime })$ is the probability
for configuration $\mathbb{A}$ to change to $\mathbb{A}^{\prime }$.
Explicitly, $\mathcal{R}$ is \cite{LiuJoladSchZia13} 
\begin{equation}
\sum\limits_{i,j\neq i}\frac{\Pi }{N}\left[ \frac{\Theta }{k_{i}}\left(
1-A_{ij}^{\prime }\right) A_{ij}+\frac{1-\Theta }{N-1-k_{i}}A_{ij}^{\prime
}\left( 1-A_{ij}\right) \right]  \label{Rs}
\end{equation}%
where $\Pi \equiv \Pi _{k\ell \neq ij}\delta \left( A_{k\ell }^{\prime
},A_{k\ell }\right) $ ensures that only $A_{ij}$ changes ($\delta $ being
the Kronecker delta) and 
\begin{equation}
\Theta \equiv \left\{ 
\begin{array}{ccc}
1 & \text{if} & k_{i}\geq \kappa \left( i\right) \\ 
0 & \text{if} & k_{i}< \kappa \left( i\right)%
\end{array}%
\right.  \label{Heavy}
\end{equation}%
is the Heaviside function, modelling the `rigid' adding/cutting behavior.

Given the explicit rates, the entire stochastic process is specified. Since
these rates are \textit{not} based on some physical process governed by a
Hamiltonian, it is important to ask if they satisfy detailed balance or not.
If they do, then the stationary distribution\footnote{%
For a physical system in thermal equilibrium, this would be the standard
Boltzmann factor.}, $\mathcal{P}^{\ast }$, can be readily constructed.
Otherwise, finding $\mathcal{P}^{\ast }$ is, in general, highly non-trivial 
\cite{Hill66}. In our case, with neither Hamiltonian nor temperature,
detailed balance can be checked via the Kolmogorov criterion \cite%
{Kolmogorov36}: A set of $R$s satisfies detailed balance if and only if the
product of $R$s around \textit{any} closed loop in configuration space is
equal to that around the reversed loop. Since it is easy to check that this
is not satisfied for some `elementary closed loops' (i.e., the operation of
adding a link $a$, then a link $b$, then deleting $a$, then deleting $b$) 
\cite{LiuJoladSchZia13}, we conclude that despite its apparent simplicity,
the model is non-trivial. A significant implication of detailed balance
violation is the presence of non-trivial, stationary probability currents, 
$\mathcal{K}^{\ast }$, even in the stationary state. As pointed out earlier 
\cite{ZS2007}, while the (time-independent) properties of an equilibrium 
system are completely specified by the steady state distribution 
$\mathcal{P}^{\ast}$, we need to specify the pair 
$\left( \mathcal{P}^{\ast },\mathcal{K}^{\ast }\right) $ 
to describe a \textit{non-equilibrium} steady state. In
Appendix A, we work out in detail the case of a small network with just four
nodes, mainly for pedagogical purposes: examining every elementary loop,
showing that many do not satisfy the Kolmogorov condition, and determining 
$\left( \mathcal{P}^{\ast },\mathcal{K}^{\ast }\right) $ explicitly.

On the other hand, it is simple to perform Monte Carlo simulations and to
discover possibly interesting phenomena. For many of the quantities we
study, the qualitative trends can be easily guessed. Apart from such
`mundane' results, some of these models do produce surprising collective
behavior. Reliable approximation schemes have been devised, so that some of
these less obvious behaviors can also be understood reasonably well. We
discuss these below.

\subsection{\textit{Homogeneous population}}

While any realistic population will display a distribution of $\kappa $s, it
is reasonable to begin with a baseline study -- a system in which every
agent shares the same $\kappa $. In ref. \cite{LiuJoladSchZia13}, we showed
that, due to the violation of detailed balance, the stationary distribution $%
\mathcal{P}^{\ast }\left( \mathbb{A}\right) $ cannot be easily found.
Nevertheless, every expectation is that the average degree is near $\kappa $%
. Indeed, given the randomness and homogeneity in the system, a naive guess
might be that the stationary ensemble is just the Erd\H{o}s-R\'{e}nyi graphs 
\cite{E-R}. In that case, the degree distribution, $\rho \left(k\right) $,
is well known: binomial (i.e., Gaussian or Poisson, in the appropriate limit
of large $N$).

\begin{figure*}[tbp]
\centering
\mbox{
    \subfigure{\includegraphics[width=3in]{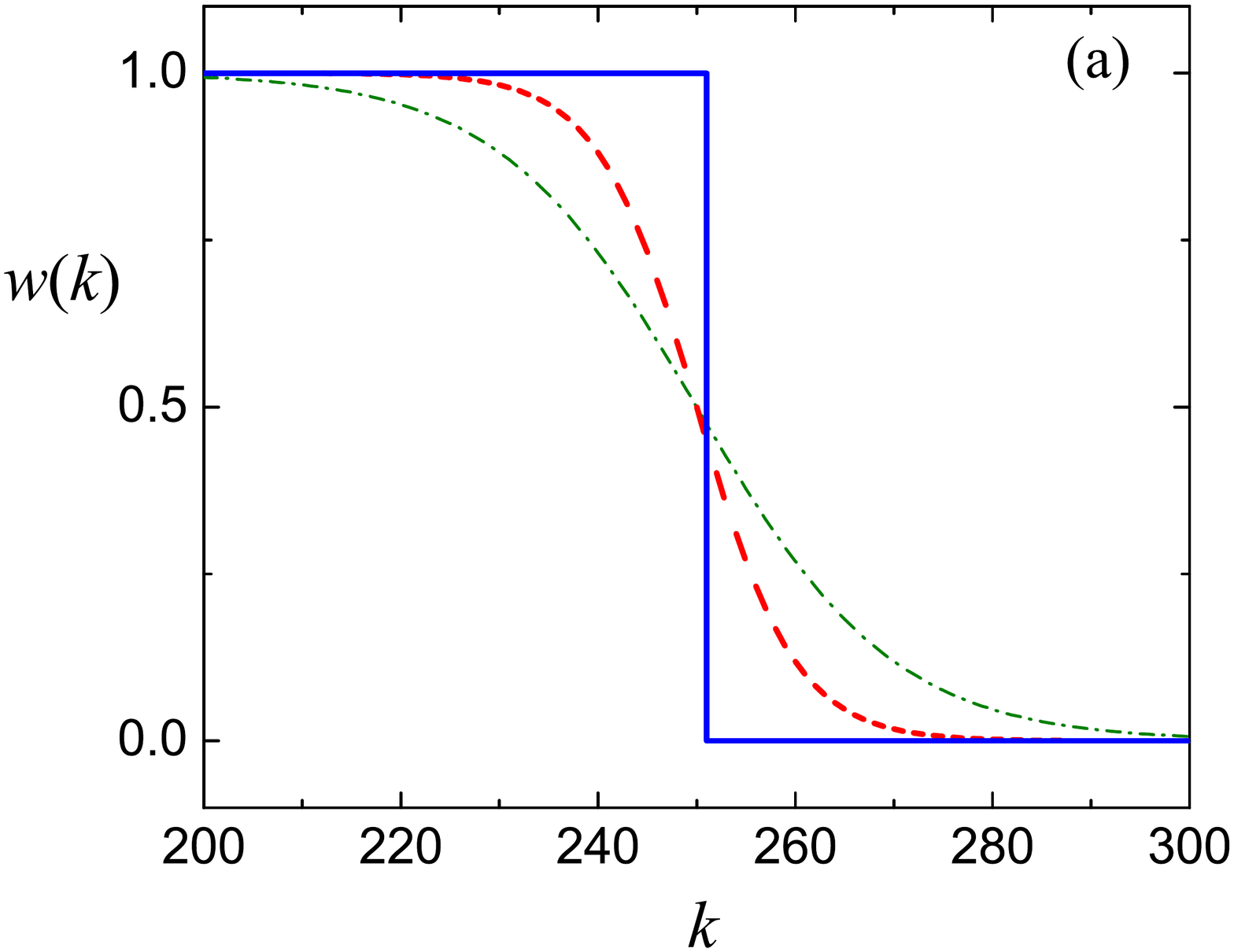}}\quad
    \subfigure{\includegraphics[width=3in]{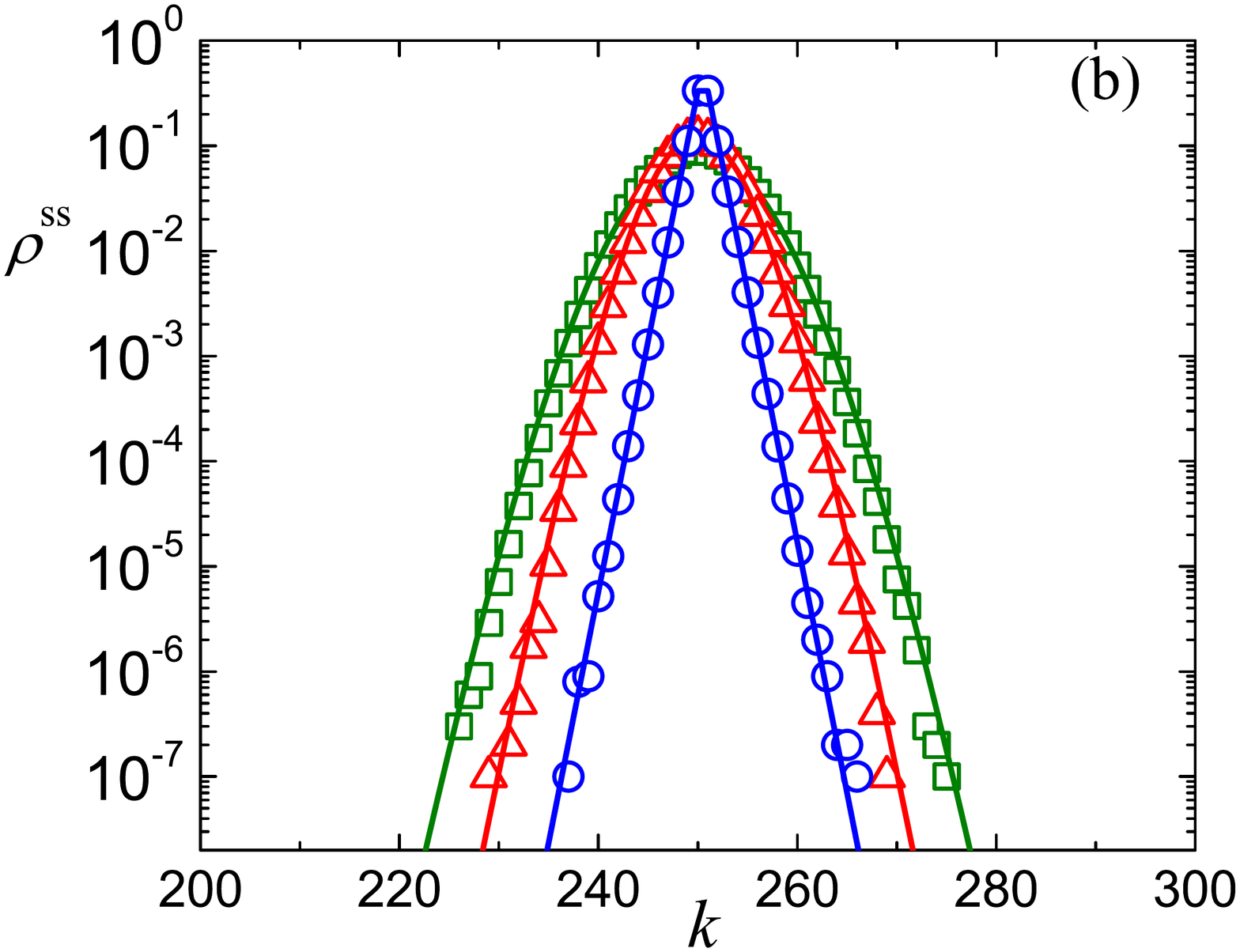}}
     }
\caption{(a) Defining $w_{\pm }$ as $w(k)$ and 1-$w(k)$, we use the form $%
w(k)=(1+e^{-\protect\beta\protect\kappa})/(1+e^{\protect\beta(k-\protect%
\kappa)})$ with $\protect\kappa=250$ in three examples: green dash-dotted
line, red dashed line and blue solid line associated with $\protect\beta=0.1$%
, $0.2$ and $\infty$ respectively. (b) The data points represent the
corresponding degree distributions of a system with $N=1000$. The solid
lines are theoretical predictions. 
(reproduced from Ref. \cite{LiuJoladSchZia13}). }
\label{homoDD}
\end{figure*}

To our surprise, from simulating a typical case with $N=1000$ and $\kappa
=250$, this system displays, to an excellent approximation, a \textit{%
Laplacian} distribution: $\rho (k)\propto \exp \left[ -\left\vert k-\kappa
\right\vert \ln 3\right] $. As illustrated in Fig. \ref{homoDD}, if we
`soften' the updating rules to the less rigid $w_{\pm }$, we find a more
rounded peak at $\kappa $ but the exponential tails persist. These features
can be understood through a simple argument: Consider an agent with $%
k>\kappa $ and the probability that its degree will increase or decrease by
one in an attempt. The former will happen only if one of the other agents
adds a link (to it). In the steady state, let us assume the partner is
equally likely to have too many or too few links. So, the probability for
adding is $1/2$. For our individual to lose a link, the argument for this $%
1/2$ also applies. But, if it is chosen, it will cut for sure. Balancing the
rates for an agent with $k$ links to gain or lose one, we are led to the
rough estimate%
\begin{equation}
\frac{1}{2}\rho \left( k\right) \thicksim \left[ \frac{1}{2}+1\right] \rho
\left( k+1\right)  \label{homoRB}
\end{equation}%
A similar argument for $k<\kappa $ leads us to $\rho (k)\propto \left(
1/3\right) ^{\left\vert k-\kappa \right\vert }$. In the limit of large 
$\kappa ,N$ with fixed $\kappa /N<1$, this argument becomes exact.

To understand simulation data with less rigid rules (e.g., 
Fig. \ref{homoDD} ), our argument can be repeated. 
In this case, the coefficients above becomes $1/2+w_{+}\left(
k;\kappa \right) $ and $1/2+w_{-}\left( k+1;\kappa \right) $, and the
agreement with data is reasonably good, as illustrated in the figure.
Needless to say, as $\kappa $
nears the two extremes -- $0$ and $N$, the distribution will be distorted
from a Laplacian.

\subsection{Two interacting communities: introverts and extroverts}

Proceeding to heterogeneous populations, the next simplest step is to
introduce two subgroups (or communities, labelled by $\alpha =1,2$ or $I,E$)
with various distinguishing properties \cite{LiuSchZia14}. Obviously, the
parameter space becomes 4-dimensional $\left( N_{\alpha },\kappa _{\alpha
}\right) $. With typically unequal $\kappa $s, we label the the group with
smaller(larger) $\kappa $ introverts(extroverts). To model \textit{%
interactions} between the communities, there are not only many possibilities
but also subtle and complex issues. In this review, we will focus mainly on
two variations: the generic ($GIE$) and the extreme ($XIE$). Though the
former is more realistic, the latter is more tractable analytically and will
be the focus of the rest of this article. As will be shown, even when
restricted to $XIE$, drastically different collective behavior emerge when
different actions on the links are introduced. In the following sections, 
we will study three variants. To distinguish them from the proto model, 
we will refer to them as Blind, Egalitarian, and Elitist XIE models, 
labelled respectively by $XIE_{bl}$, $XIE_{egal}$, and $XIE_{elit}$. 

\begin{figure*}[tbp]
\centering
\mbox{
    \subfigure{\includegraphics[width=3in]{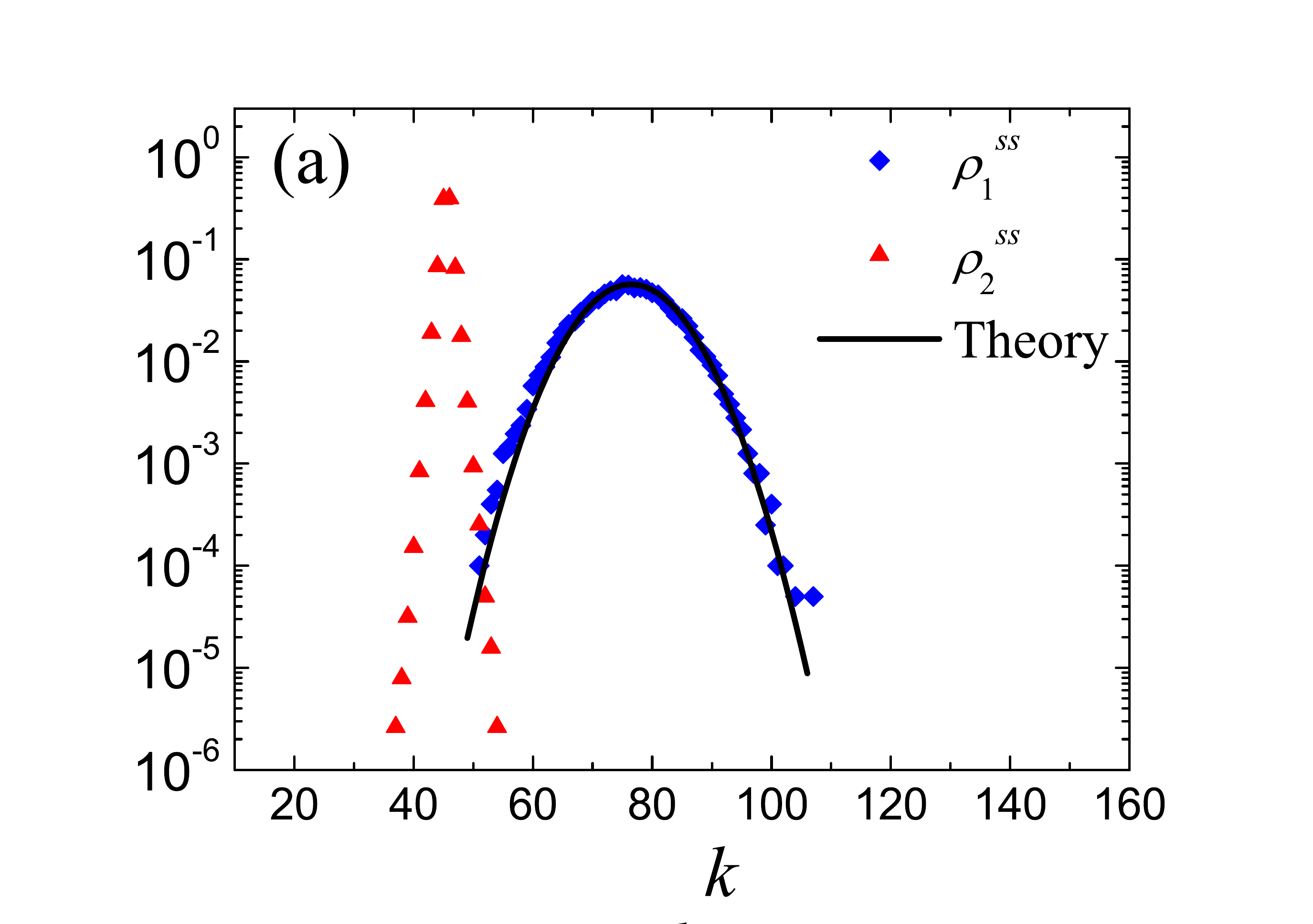}}\quad
    \subfigure{\includegraphics[width=3in]{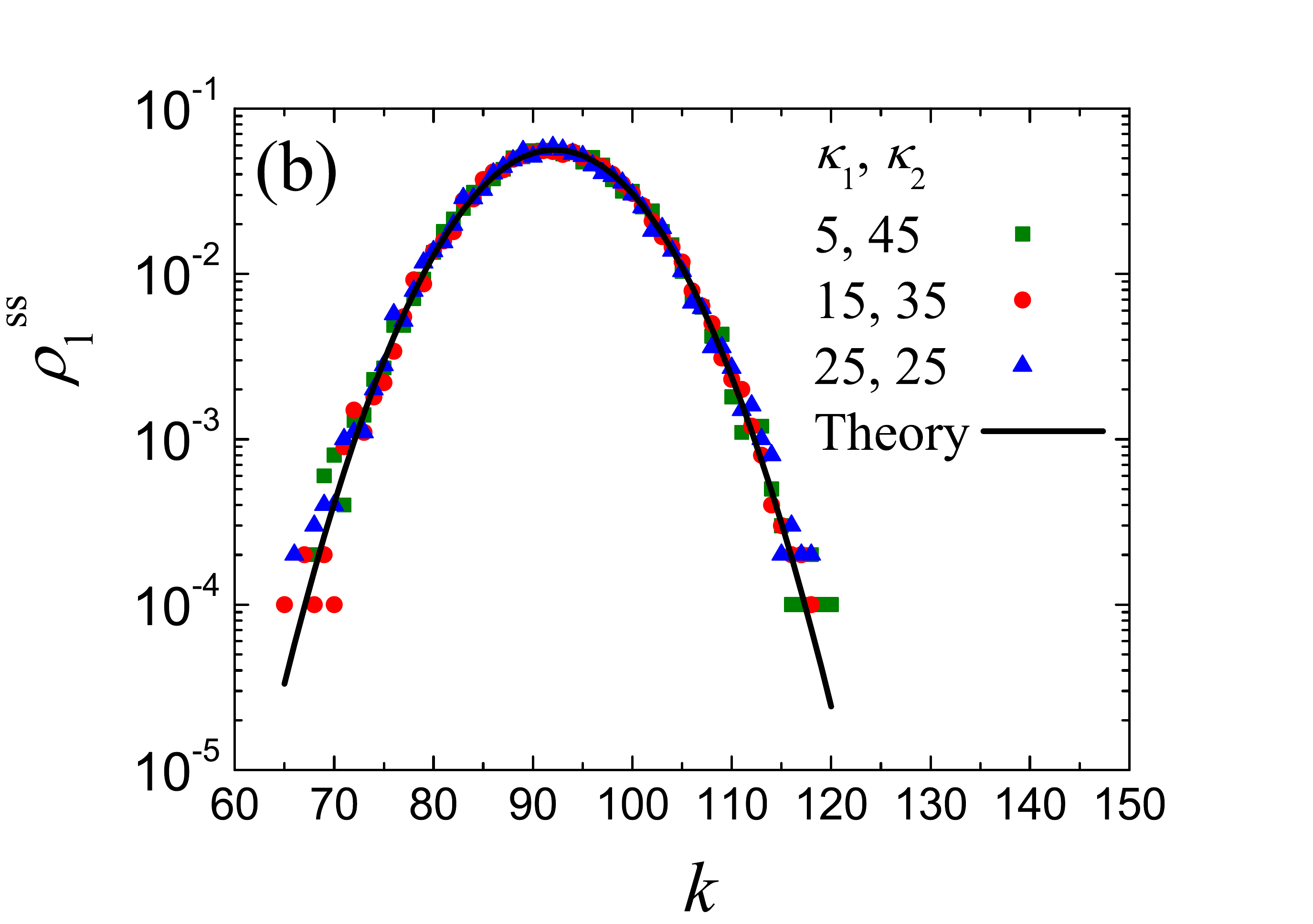}}
     }
\caption{(a) The total degree distributions for a two-network model ($%
N_{1}=10$, $N_{2}=190$, $\protect\kappa _{1}=5$, $\protect\kappa _{2}=45$
and $\protect\chi _{\protect\alpha }=0.5$): $\protect\rho _{1}^{ss}$ (blue
diamonds) and $\protect\rho _{2}^{ss}$ (red triangles). (b) The total degree
distributions $\protect\rho _{1}^{ss}$ for a two-network model with three
different values of $\protect\kappa _{1}$, $\protect\kappa _{2}$. The other
associated parameters are $N_{1}=5$, $N_{2}=195$, and $\protect\chi _{%
\protect\alpha }=0.5$. In both figures, solid lines represent theoretical
predictions. (reproduced from \cite{LiuSchZia14}) }
\label{N1<<N2}
\end{figure*}

Turning to the generic $GIE$ model with two distinct communities, it is
natural to specify an individual's preference for cutting/adding a \textit{%
intra}-group link \textit{vs.} a cross-link (a link to a partner in the
other group). The simplest way is to define $\chi _{\alpha }$, the
probability that an agent in subgroup $\alpha $ will choose a cross-link for
action. To illustrate, suppose an introvert with degree larger than $\kappa
_{I}$ is chosen, then it will cut a random cross-link with probability $\chi
_{I}$ or, with probability $1-\chi _{I}$, a random link to individuals
within the group. Clearly, $\chi $ represents how likely an agent interacts
with a member from the other community. Thus, a $\chi _{a}=0$ system
(initialized with $\mathbb{A}_{0}=0$) breaks up into two non-interacting,
homogeneous networks. The $\chi _{a}=1$ limit here is also interesting, as
the intra-links remain absent, so that only bipartite graphs are present and
the adjacency matrix reduces to the smaller $N_1 \times N_2$ -dimensional
incidence matrix $\mathbb{N}$. In either limit, a non-vanishing $\mathbb{A}%
_{0}$ can play a significant role, since the inactive sector(s) of $\mathbb{A%
}$ are equivalent to decreasing the effective $\kappa $s of agents by
different amounts. Due to such complications, we have not explored these
limits so far. Instead, we study several generic points in the 6-dimensional
space $\left( N_{\alpha },\kappa _{\alpha },\chi _{\alpha }\right) $. 
Using Monte Carlo simulations, we find certain expected properties with
typical $\chi $s. The degree distributions of each subgroup are similar to
those of homogeneous populations: Laplacians around their own $\kappa $s.
Exceptions appear when there is a large level of `frustration' \cite%
{LiuSchZia14}, i.e., when neither community can maintain link numbers close
to their preferred $\kappa $s. For example, with $\chi _{\alpha }=1/2$ say,
the extroverts will find themselves creating many links with the introverts
(roughly $N_{E}\kappa _{E}\chi _{E}$). If $N_{I}\kappa _{I}$ is much less
than this value, the introverts will struggle to cut these links, leading to
serious `frustration.' To illustrate, consider a population with $\kappa
_{I,E}=5,45$ and $N_{I,E}=10,190$. The extroverts will attempt to reach 45
links each, with about half of them ($\sim $ 4000) directed towards just 10
introverts. Not surprisingly, while the extrovert majority are content, the
few introverts will be overwhelmed and cannot achieve their preferred level
of just 5 contacts. This dichotomy is well illustrated in Fig. \ref{N1<<N2}a
. Here, $\rho _{E}$ (red triangle, $\rho
_{2}^{ss}$ in figure) displays the familiar Laplacian around $\kappa _{E}=45$%
, but $\rho _{I}$ (blue diamond, $\rho _{1}^{ss}$ in figure) is closer to a
Gaussian, peaked far beyond $\kappa _{I}=5$! Remarkably, the latter depends
mostly on the fact that $N_{1}\ll N_{2}$ and fairly independent of the
details of the $\kappa $s: Fig. \ref{N1<<N2}b .
This unexpected `universal' feature can be explained by
solving a rate balance equation in the spirit of (\ref{homoRB}) above, $\rho
_{I}\left( k\right) R\left( k\rightarrow k+1\right) =\rho _{I}\left(
k-1\right) R\left( k+1\rightarrow k\right) $, and making reasonable
arguments to arrive at the transition probabilities ($R$s) \cite{LiuSchZia14}%
.

\begin{figure}[tbp]
\centering
\includegraphics[width=3.5in]{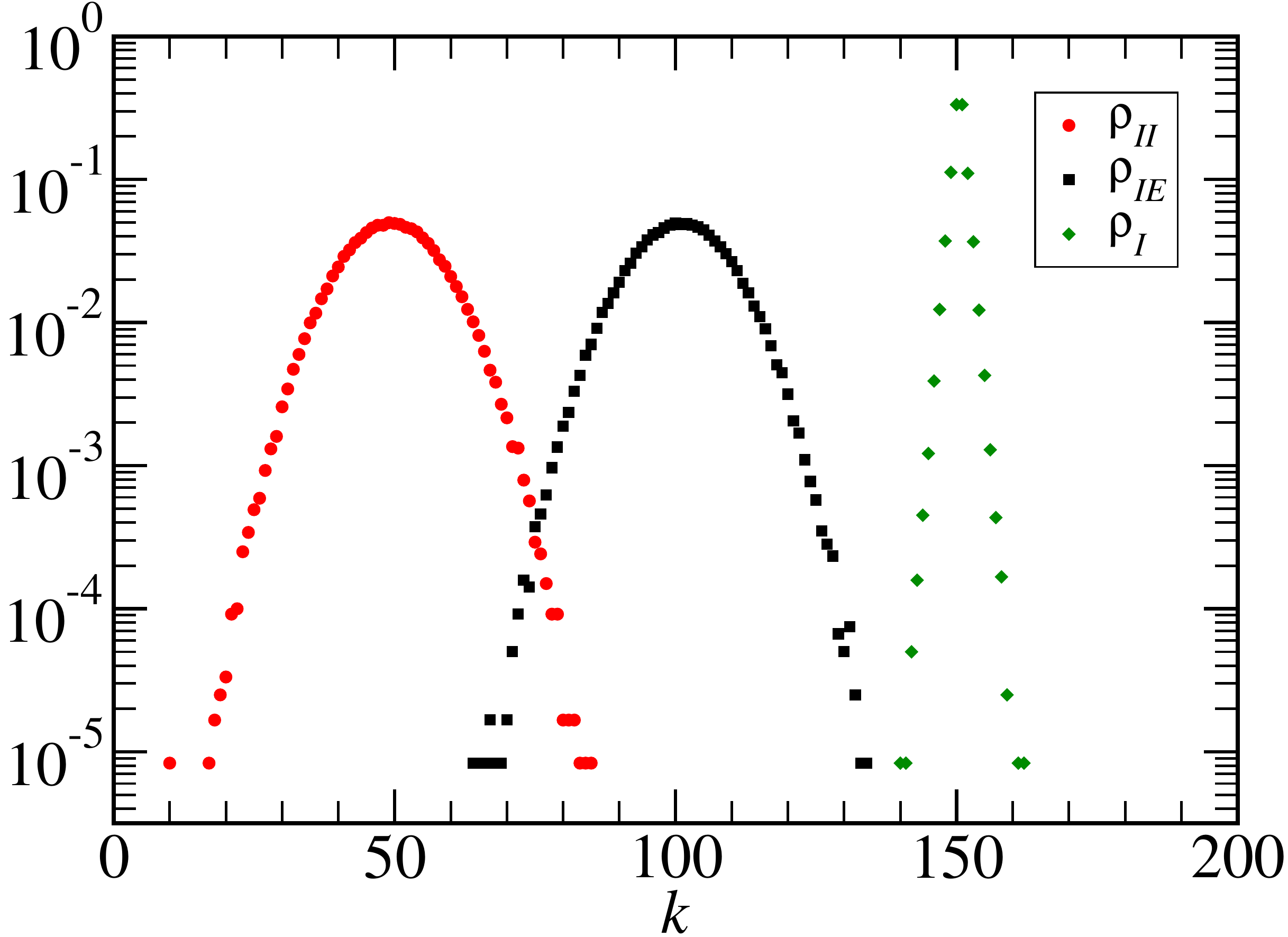}\newline
\includegraphics[width=3.5in]{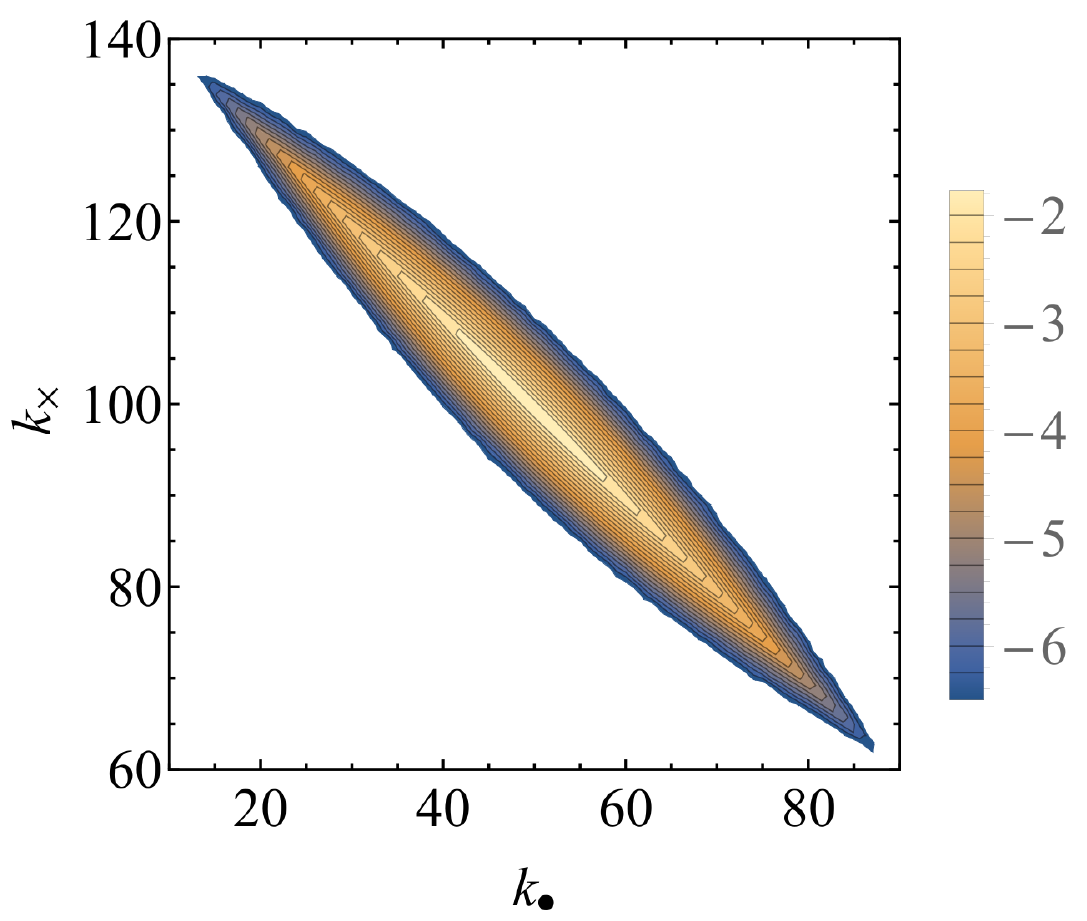}
\caption{Top figure: The degree distributions for the introverts in a model
with $N_{I}=N_{E}=1000$, $\protect\kappa _{I}=150$, $\protect\kappa _{E}=250$
and $\protect\chi _{\protect\alpha }=0.5$). The three data sets correspond
to intra-links $\protect\rho _{II}$ (red circle), cross-links $\protect\rho %
_{IE}$ (black circle), and all links $\protect\rho _{I}$ (green circle).
Bottom figure: Contour plot of the joint distribution, $\log _{10}\protect%
\rho _{I}\left( k_{\bullet },k_{\times }\right) $, showing the narrow sharp
profile when viewed along $k_{\bullet }+k_{\times }=150$, and the wide
Gaussians when projected onto either axis. White space indicates a
probability of $<10^{-6.5}$.}
\label{threeIs}
\end{figure}

So far we have focused on the ordinary degree distributions ($\rho _{I,E}$)
of the two populations. When we extend our investigations to the
distributions of inter- and intra-links ($\rho _{IE},\rho _{II},\rho
_{EI},\rho _{EE}$), we find that they are essentially broad Gaussians, even
when $\rho _{I,E}$ are narrow Laplacians. We illustrate this contrast for 
the introverts in the upper panel of Fig. \ref{threeIs}
\footnote{%
Starting with complete, empty, or random networks, we find these 
distributions after $N$ Monte Carlo steps ($N^2$ updates). They appear 
to be stable when run for a hundred times longer. That said, these 
states turn out to be quasi-stationary, as discussed later.}.
This unusual combination can be understood by studying the 
\textit{joint} distributions $\rho _{\alpha }\left( k_{\bullet },k_{\times
}\right) $, the probability that an $\alpha $ agent has $k_{\bullet }$
intra-community links and $k_{\times }$ cross-links. These are related to
the above $\rho $s by appropriate projections, e.g., $\rho _{I}\left(
k\right) =\Sigma _{k_{\bullet },k_{\times }}\delta \left( k_{\bullet
}+k_{\times },k\right) \rho _{I}\left( k_{\bullet },k_{\times }\right) $ and 
$\rho _{IE}\left( k\right) =\Sigma _{k_{\bullet }}\rho _{I}\left( k_{\bullet
},k_{\times }\right) $. Visualizing $\ln \rho _{\alpha }\left( k_{\bullet
},k_{\times }\right) $ as a landscape (Fig. \ref{threeIs}, lower panel), we
see that it resembles a relatively flat half-disk (thickness much less than
diameter), upended and laid along the diagonal $k_{\bullet }+k_{\times
}=\kappa _{\alpha }$. Then its profile along the diagonal appears sharp and
narrow, while its projection onto either axis appears broad and rounded.
However, though this scenario is adequate, we have yet to develop
quantitative explanations for the partitions between intra- and cross-links
(e.g., $150=50+100$ for the $I$s in Fig. \ref{threeIs}a) and the widths of
the Gaussians.


Finally, we discuss one of the more unexpected features: the appearance of
very large relaxation times in the system. Consider the most symmetric of
all possibilities: equal $\left( N_{\alpha },\kappa _{\alpha }\right) $s and 
$\chi _{\alpha }=0.5$. Indeed, one might have guessed that this system is
similar to a homogeneous population with $N=2N_{\alpha }$. Yet, dramatic
differences emerge. In particular, for systems with $N_{\alpha }=1000$ and $%
\kappa _{\alpha }=250$, though cross-link distributions settle into
Gaussians quite rapidly (starting with a null graph), the whole distribution
wanders very slowly (while mostly maintaining the shape). Long simulations
are required in order to study the steady state distributions in such
systems \cite{LiuJoladSchZia13}. For convenience, we resort to much smaller
networks ($N_{\alpha }=100,\kappa _{\alpha }=25$) in order to explore their
properties systematically. To characterize this seemingly bizarre behavior,
we chose to focus on a single, macroscopic quantity, $X$, the total number
of cross-links between the communities. With runs up to $10^{7}$ Monte Carlo
steps, we find that $X$ traverses much of its expected range (up to about $%
2500=N_{\alpha }\kappa _{\alpha }$), often enough for us to compile a
reliable histogram from its time trace $X\left( t\right) $. Of course, when
normalized, this histogram provides us with $P\left( X\right) $, the
probability distribution for $X$. Both this $P$ and the power spectrum
associated with $X\left( t\right) $ displays quantitatively the anomalous
large scale wanderings observed \cite{LiuSchZia14}. We found a broad plateau
in $P$ for about a third of the interval $\left[ 0,2500\right] $, (i.e.,
standard deviation of $\thicksim 300$), as well as a power spectrum that is
consistent with that of a pure random walk, $1/\omega ^{2}$, down to a lower
cut-off which corresponds to a characteristic time for a random walker to
traverse the available range of $X$. By contrast, consider a homogeneous
population with $N=200,\kappa =25$, and arbitrarily label half of them white
and the rest black. Carrying out the simulations as before, we find the
statistical properties of the total number of W-B links to be entirely
`normal.' Here, the $P$ is Gaussian, peaked around $1250=\left( N/2\right)
\kappa $ with small standard deviation ($\thicksim 25$) which is
understandable through simple arguments. Meanwhile, the power spectrum,
being essentially constant, is consistent with that of white noise. 
Let us emphasize that difference in the rules between these two models is
seemingly trivial: In the latter model, a W agent with $k>25$ chooses to cut
a random link, regardless of its being a W-B or W-W link. Thus, the two sets
of links are correlated, in the following sense. Suppose there is a
preponderance of W-B links, then it is more likely for one of those to be
cut. In the interacting communities model, however, an agent first chooses,
with equal probability, which set of links to take action. As a result, the
correlation is lost, so that each set is equally likely to lose a link,
irrespective of the preponderance of one set of links. We need a better
understanding for how the drastic differences in the collective behavior can
emerge from such a minor changes in the rules. This need also drove us to
explore simpler, limiting cases which may contain similar issues. One of
these limits is an extreme scenario -- the XIE model. 

Before discussing the special limit, we end this subsection with some
generalizations. 
Since these variants are mentioned only briefly here, we will not provide
them with special names. 
Though simplest, arguably, the rule with the $\chi $s does not represent the
most natural human behavior. For example, if we imagine two groups with
opposing political views (e.g., right/left) and that they are rather
intolerant, then a model with $\chi _{\alpha }\ll 1$ is not adequate: For
example, when a right-winger is chosen to cut a link, it is much more likely
for a cross-link to be cut. To model this kind of `dislike', we would need
four probabilities, $\chi _{\alpha ,+}\ll 1$ for adding links and $\chi
_{\alpha ,-}\simeq 1$ for cutting. Another more realistic model is to
introduce two sets of $\kappa $s: $\kappa _{\alpha \beta }$ which controls
an $\alpha $ agent's preference for links to individuals in community $\beta 
$. Thus, someone in group $1$ would prefer to have $\kappa _{11}$, $\kappa
_{12}$ intra- and inter-group links, respectively. A similar model is
specified by preferred ratios of intra- \textit{vs. } inter-group links,
modelling an agent who is most comfortable with having, say, 3/4 of its
contacts within the group and 1/4 to `strangers.' Many simulations on these
variations have been performed \cite{WJLthesis}, with most of the results
being intuitively understandable and reasonably well described by rate
balance relations like (\ref{homoRB}). Since the dynamics in both of these
models involves preferred `points' in the space of intra- and inter-link
degrees, the large scale wanderings found above are absent.

\subsection{$XIE$ - a model of extreme introverts and extroverts}

In an effort to gain some insight into the most surprising feature in the
two communities model above (giant fluctuations and large scale wandering of 
$X$), we were led to ask if this phenomenon exists in the model at an
extreme limit: $\kappa _{I}=0,\kappa _{E}=\infty $. Here, an
introvert/extrovert will always cut/add a link whenever possible. Certain
simplifications are immediately clear: In the steady state, there will be no
links in the $I$-$I$ sector, while all links will be present in the $E$-$E$
sector. The only active links will be the cross-links. Thus, the smaller
incidence matrix $\mathbb{N}$ can be used to characterize the network
instead of the larger adjacency matrix $\mathbb{A}$. An element of $\mathbb{N%
}$, denoted by $n_{ij}$, is $1$ if the link between introvert $i$ and
extrovert $j$ is present and $0$ otherwise. Thus, the $XIE$ configuration
space $\left\{ \mathbb{N} \right\} $ is identical to that of an Ising model
(on a square lattice) with%
\begin{equation}
\mathcal{N}\equiv N_{I}N_{E}
\end{equation}%
spins. Further, given that no more than one $n_{ij}$ is `flipped' in an
attempt, its evolution resembles that in the kinetic Ising model with
single-spin-flip dynamics. The major difference is, given fixed $\kappa $s,
that the only control parameters here are $\left( N_{I},N_{E}\right) $.
Meanwhile, since the probability for $n$ to `flip' from $1$ to $0$ is
associated with an $I$ being chosen, i.e., $N_{I}/N $, and similarly, $%
N_{E}/N$ for the reverse `flip,' it is natural for the variables%
\begin{equation}
\Delta \equiv N_{I}-N_{E};~~h\equiv -\frac{\Delta }{N}
\end{equation}%
to play the role of magnetic field \footnote{%
This choice of signs is for later convenience, when we focus on systems with
majority of introvert and $\Delta >0$.}. Now, it is clear that $%
X=\Sigma_{ij}n_{ij}$ here plays the role of the total magnetisation in the
Ising model: $M=2X-\mathcal{N}$. Thus, our interest here -- how does the
average $\left\langle X\right\rangle $ depend on $\left( N_{I},N_{E}\right) $
-- corresponds to the Ising equation of state: $m\equiv \left\langle
M\right\rangle /\mathcal{N}$ as a function of $\left( h,T;\mathcal{N}\right) 
$. There is no obvious $T$ in our model, and we will return to this question
later.

To reiterate, the rules of $XIE$ are minimal indeed: Choose a random
individual; if it is an $I$, cut one of its links at random; if it is an $E$%
, add a link to a random partner not already connected (to it). Most
remarkably, detailed balance is restored in this limit, so that the
stationary state can be regarded as an ordinary equilibrium one and the
distribution $\mathcal{P}^{\ast }$ can be easily determined. The result is 
\cite{CSP25,LiuSchmittmannZia12}: 
\begin{equation}
\mathcal{P}^{\ast }\left( \mathbb{N}\right) =\frac{1}{\Omega }%
\prod\limits_{i=1}^{N_{I}}\left( k_{i}!\right)
\prod\limits_{j=1}^{N_{E}}\left( p_{j}!\right)  \label{P*}
\end{equation}%
where $\Omega =\Sigma _{\mathbb{N}}\Pi \left( k_{i}!\right) \Pi \left(
p_{j}!\right) $ is a `partition function,' $k_{i}\equiv \Sigma _{j}n_{ij}$
is the degree of $i$, and $p_{j}\equiv \Sigma _{i}\bar{n}_{ij}$ ($\bar{n}%
_{ij}\equiv 1-n_{ij}$) is the complement of $q_{j}$, the degree of $j$. We
further note that the Ising symmetry has an analogue here: $%
n_{ij}\Leftrightarrow \bar{n}_{ji}$ $\oplus $ $N_{I}\Leftrightarrow N_{E}$ ,
referred to as `particle-hole' symmetry (PHS), which $\mathcal{P}^{\ast
}\left( \mathbb{N}\right) $ clearly respects. Interpreting $\mathcal{P}$ as
a Boltzmann factor, we find an explicit Hamiltonian 
\begin{equation}
\mathcal{H}\left( \mathbb{N}\right) =-\left\{ \sum\limits_{i=1}^{N_{I}}\ln
\left( \sum\limits_{j=1}^{N_{E}}n_{ij}\right) !+\sum\limits_{j=1}^{N_{E}}\ln
\left( \sum\limits_{i=1}^{N_{I}}\bar{n}_{ij}\right) !\right\}  \label{H}
\end{equation}%
(with $k_{B}T=1$). Note that this Ising Hamiltonian depends on the row- and
column sums of $\mathbb{N}$. Thus, each `spin' is coupled to all other
`spins' \textit{in its row and column}, via all types of `multi-spin'
interactions. Though the appearance of long range interactions may suggest
challenging analyses, we recall that, if an Ising Hamiltonian is only a
function of the sum over all spins, the standard mean-field approximation
becomes exact in the large-$N$ limit. The case here is more complicated, as
there are $(N_{I}+N_{E})$ different sums, but we may expect a similar
simplification in the large-$N$ limit. This expectation is realised, as we
will see in the next section.

From here, we can follow standard techniques of equilibrium statistical
mechanics to find averages ($\left\langle O\right\rangle \equiv \Sigma _{%
\mathbb{N}}O\mathcal{P}^{\ast }\left( \mathbb{N}\right) $) of various
observables, e.g., $\left\langle X\right\rangle $, $P\left( X\right)
=\left\langle \delta \left( X,\Sigma _{ij}n_{ij}\right) \right\rangle $ and $%
\rho _{I}\left( k\right) =\left\langle \delta \left( k,\Sigma
_{j}n_{ij}\right) \right\rangle $, etc. In particular, we need to compute $%
2\left\langle X\right\rangle /\mathcal{N}-1$ to find our equation of state, $%
m\left( h,\mathcal{N}\right) $. Of course, such tasks are highly
non-trivial, even when $\mathcal{P}^{\ast }$ is explicitly given. Instead,
we find that a mean-field like approximation scheme provides a good
description. For example, replacing every $n_{ij}$ by $x\equiv \left\langle
X\right\rangle /\mathcal{N}$ in $\mathcal{P}^{\ast }$, we can compute $%
P^{MFA}\left(X\right) $ for any $\left( h,\mathcal{N}\right) $. In \cite%
{LiuSchmittmannZia12}, we presented $m\left( h\right) $ using simulation
data for various systems with $N=200$, as well as $m^{MFA}$ computed from
the value of $X$ at peak of $P^{MFA}$. The two agree rather well, \textit{%
qualitatively}. The most striking feature is the hint of an extreme Thouless
effect for large $N$ \cite{BarMukamel14,BarMukamel14a}, namely, $\left\vert
m\right\vert $ is close to unity for non-vanishing $h$, so that it jumps
from almost one extreme to the other, as $h$ crosses the `critical' value of 
$0$. (Note that symmetry alone dictates $m\left( 0\right) =0$.) Unlike
typical first order transitions, e.g., Ising below $T_{c}$, because of the
absence of spatial structures in our model there is no phase coexistence, no
metastability, and no hysteresis. Instead, at $h=0$, the fluctuations are
anomalously large -- $O\left( \mathcal{N}\right) $ instead of $O\left( \sqrt{%
\mathcal{N}}\right) $. This behavior is best illustrated by simulation data
of both $P\left( X\right) $ and the time traces of $X$ in \cite%
{LiuSchmittmannZia12,BasslerLSZ15}. There, we find a broad, flat plateau
(over 70\% of the available range of $X$ in the $N=200$ case) in $P$ and $%
X\left( t\right) $ performing an unbiased random walk in this range. These
properties can be roughly understood in terms of $P^{MFA}\left( X\right) $.
For example, to leading order in $1/\mathcal{N}$, $-\ln P^{MFA}=const.+X\ln
\left( N_{I}/N_{E}\right) $, so that (a) $X$ is forced to take its extremal
values no matter how small the two population sizes differ and (b) $X$ can
take any allowed value for the $N_{I}=N_{E}$ case. In \cite{BasslerLSZ15},
we studied various $N\in \left[ 200,3200\right] $, and found that indeed, as 
$N\rightarrow \infty $, both $1-\left\vert m\left( h=0.01\right) \right\vert 
$ and $1-\left\vert m\left( h=2/N\right) \right\vert $ approach zero.
However, the effective power laws, $N^{-0.71}$ and $N^{-0.34}$ respectively,
may indicate a large crossover regime, the full understanding of which
remains a challenge.

\begin{figure}[tbp]
\centering
\includegraphics[width=3.5in]{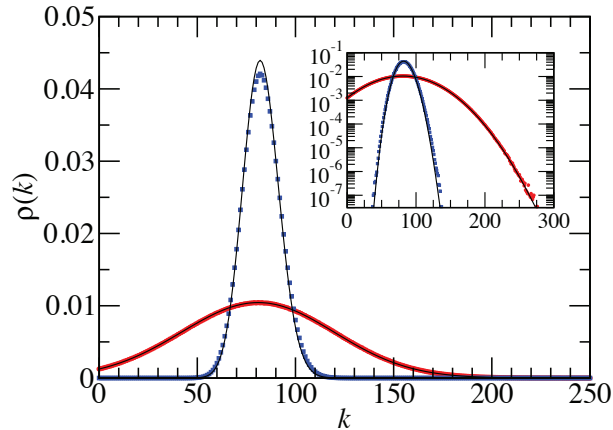}
\caption{ The degree distributions for an introvert and an extrovert in a
large, near-critical system: $N_I,N_E=1601,1599$. Insets show the same
points in log-linear plot. Theoretical predictions are shown as black lines.
Data points are circles, red and blue for the introvert and extrovert,
respectively. }
\label{3200dd}
\end{figure}

Not surprisingly, similar extreme behavior is displayed in the degree
distributions $\rho _{I,E}\left( k\right) $ \cite{BasslerLSZ15}. To
understand these, we proposed a MFA for their transition probabilities,
along the lines above. In this sense, it is an approximation on the dynamics
of the system, rather than on the static formula $\rho _{I}\left( k\right)
=\left\langle \delta \left( k,\Sigma _{j}n_{ij}\right) \right\rangle $. For $%
k>0$, $R\left( k\rightarrow k-1\right) $ for an introvert is trivially $1/N$%
, since it always cuts a link when chosen. For $R\left( k\rightarrow
k+1\right) $, we note that $N_{E}-k$ extroverts can add a link to it. But we
need the chances each will add a link to our introvert. In the spirit of
MFA, the most natural estimate for this probability is its average: $%
\left\langle \Theta \left( p\right) /p\right\rangle $, where $p$ is the
number of `holes' the extrovert has. We chose a slightly more convenient
estimate, $1/\left\langle p\right\rangle ^{\prime }$, where $\left\langle
\right\rangle ^{\prime }$ is the average over \textit{only} those extroverts
with $p>0$. Thus, we find a recursion relation for the approximate%
\begin{equation}
\tilde{\rho}_{I}\left( k+1\right) =\tilde{\rho}_{I}\left( k\right) \frac{%
R\left( k\rightarrow k+1\right) }{R\left( k+1\rightarrow k\right) } \cong 
\frac{N_{E}-k}{\left\langle p\right\rangle ^{\prime }}\tilde{\rho}_{I}\left(
k\right)  \label{RR}
\end{equation}
where the tilde reminds us that this is a MFA. Performing a similar
treatment for the extroverts, we get approximate expressions for both steady
state distributions $\tilde{\rho}_{I,E}\left( k\right) $. Now, these can be
used to compute unknowns like $\left\langle k\right\rangle ^{\prime }$ and $%
\left\langle p\right\rangle ^{\prime }$, so that a self-consistent condition
must be imposed. (In Appendix B, we provide a simpler alternative route to
these $\tilde{\rho}$s.) The result are just functions of $\left(
N_{I},N_{E}\right) $, with no free fitting parameters! They turn out to be
truncated Poisson distributions:%
\begin{equation}
\tilde{\rho}_{I}\left( k\right) =\frac{\left( \left\langle p\right\rangle
^{\prime }\right) ^{N_{E}-k}}{Z_{I}\left( N_{E}-k\right) !};~~\tilde{\zeta}%
_{E}\left( p\right) =\frac{\left( \left\langle k\right\rangle ^{\prime
}\right) ^{N_{I}-p}}{Z_{E}\left( N_{I}-p\right) !}  \label{tilde-dists}
\end{equation}%
which respect PHS manifestly. Here, the $Z$s are the normalization factors
and $\tilde{\zeta}$ is the MFA `hole' distribution of the extroverts (i.e.,
if we denote the degree of an $E$ by $q$, then its degree distribution is
given by $\tilde{\rho}_{E}\left( q\right) =\tilde{\zeta}_{E}\left(
N_{I}-p\right) $). Plotting these predictions with Monte Carlo data from
very long runs with various $N=200$ systems, we find that they are
statistically indistinguishable for all systems \cite{BasslerLSZ15} \textit{%
except} the critical case (where they fail badly). Though we expect MFA mean
field approximations to be good only `far from criticality,' we were
surprised by the excellent overall agreement here. Indeed, in an effort to
get closer to criticality, we study a much larger system $N$ ($3200 $) with
a much smaller $h$ ($1/1600$). In Fig. \ref{3200dd} we see that, while $%
\tilde{\rho}_{E}$ shows discernible deviations from data, $\tilde{\rho}_{I}$
remains a `perfect fit' down to the level of $10^{-7}$!


\begin{figure}[tbp]
\centering
\includegraphics[width=3.5in]{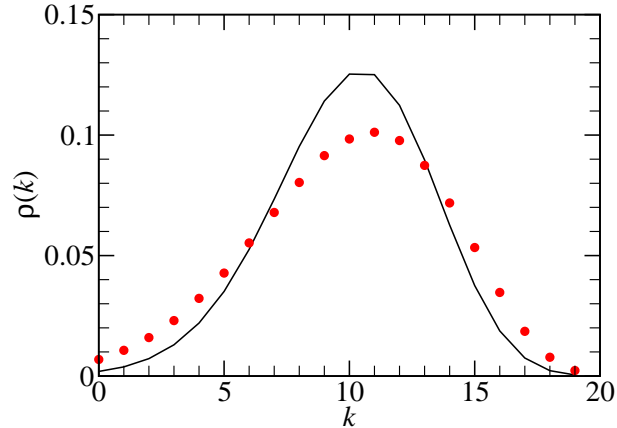}
\caption{The degree distributions for an introvert in a small, critical
system: $N_I=N_E=20$. The MFA prediction is shown as a thin black line.
The data are shown as solid red circles. }
\label{2020dd}
\end{figure}


To summarize, the $XIE$ model, despite its minimal nature, displays
intriguing and rich phenomena. The presence of an extreme Thouless effect,
i.e., 
\begin{equation}
m\left( h\right) =sign\left( h\right)
\end{equation}%
in the $N\rightarrow \infty $ limit, suggested by these studies, will be
proved in the next section. Meanwhile, a dynamic MFA for the degree
distributions appears to work quite well for off-critical ($h\neq 0$)
systems, even for systems as large as $N=3200$. On the other hand, it
performs quite poorly for the critical case, even for a small system like $%
(20,20)$ (Fig. \ref{2020dd}), providing hints that the real distributions
are much broader than expected. On the other hand, by symmetry, $%
\left\langle X\right\rangle =\mathcal{N}/2$ at criticality and so, $m\left(
h=0\right) =0$. Thus, our MFA predictions for $m\left( h\right) $ are in
excellent agreement with all our simulations so far. Clearly, there are many
interesting issue of finite-size effects, e.g., how does $m\left( h,N\right)
-sign\left( h\right) $ vanish with $1/N$?

Before turning to such recent developments, let us highlight the crucial
role of our specific dynamics in producing all the interesting phenomena
observed, by studying the same model with a seemingly the same, yet slightly
different, set of rules. We should emphasize that all our dynamic networks
are \textit{defined by} the rules, as opposed to imposing detailed balance
with respect to a given Hamiltonian (in order to arrive at the associated
Boltzmann distribution). Therefore, even if two sets of rules appear similar
and satisfy the Kolmogorov criterion, the steady state can be quite distinct.

To illustrate this important connection is part of the motivation for
studying this variant of the $XIE$. Denoted by $XIE_{bl}$ and named the
`blind' $XIE$, this model consists of agents whose actions are \textit{blind}
to the existing state of the links. Thus, when an $I$ is chosen to act, it
randomly chooses an $E$ first and then makes sure there is no link
(regardless of whether there was one or not). Similarly, an $E$ will choose
a random $I$ and ensures that a link is present. Though this dynamics is
seemingly the same, its consequences on the system are profoundly different.
Also known as the `heat-bath' dynamics in the context of Ising simulations,
these rules assigns a new configuration with probabilities \textit{%
independent} of the previous $\mathbb{N}$. As a result, $\mathcal{P}^{\ast }$
is trivially $\Pi _{ij}\left( N_{I}\delta \left( n_{ij}\right) +N_{E}\delta
\left( n_{ij},1\right) \right) /N$ and corresponds to non-interacting Ising
spins in an external field. In this case, the equation of state is exactly $%
m=h$, with neither phase transitions nor anomalous behavior \cite%
{LiuSchmittmannZia12}. 

\section{New analytical results in the large $N$ limit of the $XIE$ model}

Though it is clear that the MFA for $\tilde{\rho}$ are quite adequate for
the $N$s we explored so far, two questions stand out. Why is MFA so good? Is
the $N\rightarrow \infty $ limit analytically accessible? In this section,
we provide some answers. Before turning to the details, let us begin with
some general observations which facilitate the analysis of the large $N$
limit, as well as forming an intuitive picture of the system's unusual
behavior.

As shown above, the standard $XIE$ model displays a sharp transition when $%
N_{I}$ crosses $N_{E}$. With finite $N$, this critical point can not be
approached closer than $1/N$. This disadvantage can be remedied by
introducing a continuous, fugacity-like variable $z$. It represents a bias
in favor of the extroverts (each selected with probability $z/\left(
N_{I}+zN_{E}\right) $ to add a link) and plays the role for generating all
moments of $X$. Thus, we consider 
\begin{equation}
\mathcal{P}_{z}^{\ast }=\frac{z^{X}}{\Omega \left( z\right) }%
\prod\limits_{i=1}^{N_{I}}\left( k_{i}!\right)
\prod\limits_{j=1}^{N_{E}}\left( p_{j}!\right)
\end{equation}%
where%
\begin{equation}
\Omega (z)=\sum_{\mathbb{N}}\left[ z^{X}\prod_{i}(k_{i}!)\prod_{j}(p_{j})!%
\right]
\end{equation}%
Of course, we have the Hamiltonian ${\mathcal{H}}\left( z\right) =-\left\{
\Sigma _{i}\ln (k_{i}!)+\Sigma _{j}\ln (p_{j}!)+\ln z\Sigma
_{i,j}n_{ij}\right\} $. Note that $z$ does not add new physics to the $XIE$
model, in the sense that the essence of our system's behavior can be studied
by varying just the numbers of each subgroup. By extending parameter space
to $\left( z,N_{E},N_{I}\right) $, we will find that, in the $N\rightarrow
\infty $ limit, 
\begin{equation}
\alpha \equiv zN_{E}/N_{I}
\end{equation}%
becomes the key control variable, with $\alpha _{c}=1$ being the critical
point.

We observe that the lowest value of energy scales as $N^{2}\ln N $ for $%
N\rightarrow \infty $ with $\alpha$ fixed. This is stronger than the
quadratic dependence on the number of agents! Such non-linear increase of
`energy' with system size reflects the fact that, in the stationary state,
some configurations may occur with very low probability. Examples of this
behavior in previously studied models include the so-called ABC model\cite%
{evans} and the Oslo rice-pile model \cite{pradhan}. This non-extensivity of
the `energy' causes no real problems in our model, as all ensemble averages
are well-behaved in the large $N$ limit. Since the number of configurations
only increases as $2^{ N_{I}N_{E}}$, the `entropy term' in the effective
free energy scales as $N^{2}$. Thus, in the large $N$ limit, the behavior of
the system is dominated by the energy term. Since the empty or the complete
(bipartite) graph has the lowest energy (depending on $\alpha $), we have a
good intuitive picture for the emergence of the extreme Thouless effect,
namely, the fraction $\left\langle X\right\rangle /\mathcal{N}$ jumping from 
$0$ to $1$ as $\alpha $ crosses unity. This observation also naturally lead
to consider variational and perturbative approaches.

Even for non-extensive systems, the variational principle of minimizing free
energy allows us to find the best choice of parameters in a trial
probability measure for approximating the true $\mathcal{P}^{\ast }$.
Consider a trial state ${\mathcal{S}}_{X}$ in which all $\binom{\mathcal{N}}{%
X}$ configurations with exactly $X$ links are equally likely. Then, the
restricted partition function of the system, $\Omega (X)$, satisfies%
\begin{equation}
\ln \Omega (X)\geq \ln {\binom{\mathcal{N}}{X}}-\langle {\mathcal{H}}\rangle
_{{\mathcal{S}}_{X}}
\end{equation}%
where, in the last term, the average is over the trial state. Away from
criticality, the distributions of $k_{i}$ and $q_{j}$ are sharply peaked
around their mean values and we can use Stirling approximation for $\ln
(k!)\cong k\ln k$, a slowly varying function of $k$. Then, the terms
involving $\ln X$ cancel and the final result is a simple linear function
of $X$, with coefficient $\ln \alpha $. Apart from $z$, this alternative
approach reproduces the linear term in the mean field free energy in \cite%
{CSP25} and so, also predicts an extreme Thouless effect with $\alpha _{c}=1$%
.

\subsection{Effective Hamiltonian for introvert dominated systems ($\protect%
\alpha <1$)}

Although our system displays two phases, they are related by PHS. Without
loss of generality and for ease of presentation, we restrict our discussion
to the low density phase. Here, the typical degrees of both agents should be
small compared to $N$. Then, it is convenient to rewrite ${\mathcal{H}}
\left( z\right) $ as 
\begin{equation}
{\mathcal{H}}\left( z\right) =-\sum_{i=1}^{N_{I}}\left[ \ln (k_{i}!)+(\ln
z)k_{i}\right] -\sum_{j=1}^{N_{E}}\log \left( N_{I}-q_{j}\right) !
\label{Hz}
\end{equation}%
(where $q_{j}$ is the degree of extrovert $j$), as we seek limits of $\left(
k,q\right) /N\rightarrow 0$. Now, there are many ways to approach such a
limit, e.g., $\left( 1-\alpha \right) \propto N^{\phi -1}$ (or simply $%
\Delta \propto N^{\phi }$ for $z=1$). For the case $\phi =1$ (fixed $\alpha
<1$), we can show that the results obtained through MFA above are exact in
this limit and offer systematic corrections through a perturbative approach.
We also find a scaling representation of the degree distribution in the $%
\phi =1/2$ case, as well as some aspects of the critical $\alpha =1$ system.

While our conclusions should be valid for all $\phi \in \left[ 1/2,1\right] $%
, the analysis of the $\phi <1/2$ regime remains challenging. In particular,
the $\phi =0$ case (e.g., even $N$, $\Delta =2$, $z=1$) is of special
interest: Starting with a large and equal numbers of introverts and
extroverts, the expected fraction of cross links is $1/2$. How can letting a 
\textit{single} agent change sides have such a dramatic effect on this
fraction (e.g., 0.14 in the 101 $I$s vs. 99 $E$'s case)? Also, will the
observed power law $\left( 1-\left\vert m\right\vert \right) \thicksim
N^{-0.34}$ for $100\lesssim N\lesssim 3000$ eventually cross over to $1/3$?
or $1/2$? or perhaps some irrational value? Certainly, to understand the
quantitative aspects this behavior will not be trivial.

\subsubsection{Large $N$ limit with fixed $\protect\alpha <1$}

First, note that this condition corresponds to fixed ratios $N_{E}/N_{I}$ or 
$\Delta /N$ with $z=1$ (parameters used in all simulation data here). Here,
we expect $k,q={\mathcal{O}}\left( 1\right) $ and $x={\mathcal{O}}\left(
1/N\right) $, so that we can exploit the following approximation for terms
in the second sum in (\ref{Hz}). Writing%
\begin{equation}
\frac{N_{I}!}{(N_{I}-q)!}=\frac{N_{I}^{q}}{F(q;N_{I})}  \label{F}
\end{equation}%
where 
\begin{equation}
F(\ell ;M)=\prod_{r=1}^{\ell }\left[ 1-\frac{r-1}{M}\right] ^{-1};~~\ell >0
\label{Fdef}
\end{equation}%
and $F(0;M)\equiv 1$, we find

\begin{equation}
\ln F(\ell ;M)=-\sum_{r=1}^{\ell} \ln \left[ 1-\frac{r-1}{M}\right] \cong 
\frac{\ell \left( \ell -1\right) }{2M}+...  \label{Fapprox}
\end{equation}%
for large $M$ and $\ell \ll M$. Given $\ln F(q;N_{I})={\mathcal{O}}\left(
1/N\right) $, a natural perturbative approach emerges:%
\begin{equation}
{\mathcal{H}}\left( z\right) ={\mathcal{H}}_{0}+{\mathcal{H}}_{int}
\end{equation}%
where 
\begin{equation}
{\mathcal{H}}_{0}=-\sum_{i}\left[ \ln \left( k_{i}!\right) +k_{i}\left( \ln 
\frac{z}{N_{I}}\right) \right] -N_{E}\ln N_{I}!  \label{H0}
\end{equation}%
and%
\begin{equation}
{\mathcal{H}}_{int}=-\sum_{j}\ln F(q_{j},N_{I})  \label{Hint}
\end{equation}

We coined the term `interaction Hamiltonian' for (\ref{Hint}) since, under ${%
\mathcal{H}}_0$, our system reduces, if it were absent, to a collection of 
\textit{non-interacting} $k $s. Then, the summation over the configuration
of links attached to different introverts can be carried out independently.
As there are $\binom{N_{E}}{k_{i}}$ ways to assign $k$ links to a given
introvert, it is easy to obtain, e.g., the associated partition function: 
\begin{equation}
\Omega _{0}=(N_{I}!)^{N_{E}}[\omega _{0}]^{N_{I}}
\end{equation}%
with 
\begin{equation}
\omega _{0}=\sum_{k}\alpha ^{k}F(k,N_{E})  \label{omega0}
\end{equation}%
Since $F\rightarrow 1$ for large $N$, we find $\omega _{0}\rightarrow
1/(1-\alpha )$. Identifying $\omega _{0}$ as the partition sum over a single
introvert, we arrive at its degree distribution%
\begin{equation}
\rho _{I,0}\left( k\right) \rightarrow (1-\alpha )\alpha ^{k}
\end{equation}%
and the average $\left\langle k\right\rangle _{0}=\alpha /\left( 1-\alpha
\right) $ (which is just $N_{E}/\Delta $ when $z=1$). Here, the subscript $0$
indicate averages with $e^{-{\mathcal{H}}_{0}}$.

Turning to the extroverts' distribution, we note that the number of
connections $q_{j}$ of extrovert $j$ is a sum of $N_{I}$ independent
contributions. Any particular introvert will be connected to the given agent
is ${\mathcal{O}}\left( 1/N_{E}\right) $. Hence, it follows that the degree
distribution of an extrovert is a Poisson distribution. The mean $%
\left\langle q\right\rangle _{0}$ is clearly fixed by $\left\langle
q\right\rangle _{0}N_{E}=\left\langle X\right\rangle _{0}=\left\langle
k\right\rangle _{0}N_{I}$ (i.e., $N_{I}/\Delta $ when $z=1$), so that we find%
\begin{equation}
\rho _{E,0}\left( q\right) \rightarrow \left. e^{-\gamma }\gamma ^{q}\right/
q!
\end{equation}%
where $\gamma \equiv \left\langle q\right\rangle _{0}$. As expected, these
results are identical to appropriate limits of (\ref{tilde-dists}). For
example, $\left\langle p\right\rangle ^{\prime }\cong N_{I}-\left\langle
q\right\rangle \rightarrow N_{I}$ and $\left( N_{E}-k\right) !\rightarrow $ $%
N_{E}^{k}/N!$ so that $\tilde{\rho}_{I}\rightarrow \left( \Delta
/N_{I}\right) \left( N_{E}/N_{I}\right) ^{k}$. While the unified description
for a \textit{finite} system is a \textit{truncated }Poisson distribution,
we see that, in this thermodynamic limit ($\phi =1$), $\tilde{\rho}_{I}$ and 
$\tilde{\rho}_{E}$ approaches, respectively, an exponential a pure Poisson.
In Appendix C, we highlight the key ingredients behind the emergence of
these very different limits.

\subsubsection{Scaling regime near criticality: $\left( 1-\protect\alpha %
\right) \propto N^{-1/2}$}

While the analysis above allows us to approach the critical point as long as 
$N\gg \left( 1-\alpha \right) ^{-1}$, here we present a scaling study, valid
beyond this regime. Noting that $N_{I,E}\cong N/2$ here, it is convenient to
define the scaling variables%
\begin{equation}
\tau \equiv \left( -\ln \alpha \right) N_{I}^{1/2};~~\tilde{k}\equiv
k/N_{I}^{1/2}
\end{equation}%
From Eq. (\ref{omega0}), we see that 
\begin{equation}
\rho _{I,0}\left( k\right) \propto \frac{\alpha ^{k}}{F(k;N_{E})}.
\end{equation}%
Instead of the limit $N_{E}\rightarrow \infty ,k={\mathcal{O}}\left(
1\right) $ above, we study $k$s ranging up to ${\mathcal{O}}\left( \sqrt{N}%
\right) $ here. Using Eq. (\ref{Fapprox}), we find%
\begin{equation}
F(k;N_{E})\cong \exp (k^{2}/2N_{E})=\exp (\tilde{k}^{2}/2)
\end{equation}%
and, writing $\alpha ^{k}=e^{-\tilde{\Delta}\tilde{k}}$, we obtain $\rho
_{I,0}\propto \exp \left\{ -\tau \tilde{k}-\tilde{k}^{2}/2\right\} $ to
leading order. Thus, we arrive at the scaling form 
\begin{equation}
\rho _{I,0}\left( k;\alpha ,N_{I}\right) \thicksim N_{I}^{-1/2}\Phi \left( 
\tilde{k},\tau \right)  \label{scaling}
\end{equation}%
where $\Phi (x,\tau )\equiv \exp (-\tau x-x^{2}/2)$. While extensive
simulations are yet to be performed in this scaling regime, this prediction
is in reasonably good agreement with existing data for $\left(
N_{I},N_{E}\right) =\left( 110,90\right) $.

Meanwhile, for the $\left( 1601,1599\right) $ case shown in Fig. \ref{3200dd}%
, we see that $\rho _{I}\left( k\right) $ \textit{increases} by an order of
magnitude before decreasing monotonically. As the scaling form (\ref{scaling}%
) does not increase with $k$ (since $\tau >0$), this behavior hints at the
range of validity of this scaling regime. If we rely on the MFA in the
previous section, then Eqn. (\ref{RR}) gives an implicit condition for
monotonicity: $N_{E}\leq \left\langle p\right\rangle ^{\prime }$. Using the
results from Appendix B, the conclusion is entirely consistent with the
assumptions for this regime, namely, that this form cannot describe the data
when $1-\alpha $ drops below $\sqrt{1/N_{E}}$.

\subsection{Contributions from $\mathcal{H}_{int}$}

With this understanding of the unperturbed system, let us turn to the
perturbation. We will show that these contributions are not only small, but
also vanishes with $1/N$. Unlike the MFA above, this machinery here allows a
systematic study, so that corrections for both the $\phi =1$ and $1/2$
regimes can be computed. A detailed study is beyond the scope of this work;
only highlights will be reported here.

From Eqns. (\ref{Hint}) and (\ref{Fapprox}), we see that%
\begin{equation}
{\mathcal{H}}_{int}\cong -\frac{1}{2N_{I}}\sum_{j}q_{j}(q_{j}-1)
\end{equation}%
While this varies from one configuration to another, its fluctuations are
quite limited. Since each term in $\mathcal{H}_{int}$ is of order $1/N$, the
total sum is only $\mathcal{O}(1)$. This could be compared to $\mathcal{H}%
_{0}$, which diverges as $N^{2}\ln N$. In fact, the situation is even
better. This is a sum of many small terms, and different $q_{j}$ that appear
in the sum are weakly correlated random variables. Such a sum has even
smaller fluctuations. Here, each term is of order $1/N$ and its fluctuation
is also $\mathcal{O}(1/N)$. Thus, the sum is $\mathcal{O}(1)$ while the
associated variance, $\left\langle {\mathcal{H}}_{int}^{2}\right\rangle_{0}-%
\left\langle {\mathcal{H}}_{int}\right\rangle_{0}^{2}$ , is of $\mathcal{O}%
(1/N)$. This analysis show that, for large $N$, the effects of the
perturbation become negligible. Of course, this argument breaks down
precisely at the critical point, where the correlations between different $%
q_{j}$s are no longer small. These considerations provide the insight into
why the simulation data are so well captured by the MFA sketched in the
previous section.

\subsection{Plateau of $P\left( X\right) $ at criticality and its edges}

Finally, we turn to the extraordinary fluctuations in the critical system.
For simplicity, we focus on $z=1$ and define $L\equiv N_{I}=N_{E}$. There is
a simple way to understand the broad and flat plateau in $P\left( X\right) $
and the random walk nature of $X\left( t\right) $. Consider the `motion'
from a given $X$: It will increase by unity if an $E$ is chosen \textit{and }%
it is not connected to all $I$s. Similarly, it will decrease when a
partially connected $I$ is chosen. Otherwise, it does not `move.' Since the
probabilities for choosing either agent is $1/2$, the probabilities for $X$
to change or not are simply $\left[ 1-\zeta _{E}\left( 0|X\right) \right] /2$%
, $\left[ 1-\rho _{I}\left( 0|X\right) \right] /2$, and $\left[ \zeta
_{E}\left( 0|X\right) +\rho _{I}\left( 0|X\right) \right] /2$. Here, $|X$
means `given the number of cross links is $X$.' Now, the average degree of
any individual is just $X/L$. For large $N$ and $X$ far from either $0$ or $%
L^{2}$, $\zeta _{E}\left( 0|X\right) ,\rho _{I}\left( 0|X\right) \cong 0$ is
a good approximation, so that $X$ performs an unbiased random walk to $X\pm
1 $. However, if $X$ wanders near one of its boundaries ($0,L^{2}$), then $%
\rho _{I}\left( 0|X\right) $ or $\zeta _{E}\left( 0|X\right) $ can be
non-trivial, so that $X$ is biased to move towards the center. This argument
can be sharpened to locate the `edge' of the plateau.

Suppose $X\thicksim {\mathcal{O}}\left( L^{3/2}\right) $. From the
expression for partition function $\omega _{0}$, putting $\alpha =1$, we get
for the unperturbed Hamiltonian $\rho _{I}\left( k\right) \propto \left.
L!\right/ \left[ (L-k)!L^{k}\omega _{0}\right] \propto F(k)\approx \exp
(-k(k-1)/2L)$ . Normalizing, we have 
\begin{equation}
\rho _{I}\left( k\right) \cong \sqrt{\frac{2}{\pi L}}\exp \left\{ -\frac{%
k(k-1)}{2L}\right\}  \label{eq:p-i}
\end{equation}%
This function is approximately constant for $k\ll \sqrt{L}$, and decreases
quickly for larger $k$ . Thus, the probability that $k=0$ is of order $%
L^{-1/2}$, when $k<L^{1/2}$. In terms of the motion of the random walker,
when $X$ is of order $L^{3/2}$, each introvert agent has only ${\mathcal{O}}
(L^{1/2})$ connections on the average and there is a significant probability
that a chosen introvert has no contacts. Then $X$ will not decrease. Thus,
for $X$ $\lesssim $ $L^{3/2}$, the walker feels a net bias, of ${\mathcal{O}}%
\left( L^{-1/2}\right) $, towards larger $X$. In this sense, the motion can
be interpreted as a particle in a potential well which is nearly flat in the
range $cL^{3/2}\leq X\leq L^{2}-cL^{3/2}$, ($c$ being a constant of order $1$%
) and an approximately constant bias of ${\mathcal{O}}(L^{-1/2})$ when $%
X<cL^{3/2}$. This picture is consistent with the preliminary studies \cite%
{GreilBassler} of how the `left edge of the plateau,' $x_{edge}$ varies with 
$L$. There, the effective exponent is also decreasing, so that $%
x_{edge}\thicksim L^{-.38}$ for $L=1778$. As in the previous paragraphs, we
believe that finite size corrections are non-trivial, even when $L\thicksim
2000$. As a result, we may need to run with much larger systems to check if
the exponent does converge to its asymptotic value $1/2$.

\section{Variants of $XIE$ with preferential attachment}

In the $XIE$ model, an introvert chooses a \textit{random} link to cut,
while in the Blind-$XIE$, it chooses a random \textit{partner} (and cuts the
link if present). Many of us are more selective when we face choices. Thus,
we consider two other variants, modelling more discerning human behavior.
Again, for simplicity, we introduce extreme versions to study, in this case,
with preferential attachment and detachment. As may be expected,
dramatically different phenomena emerge. Unlike the models above, the
dynamics of these $XIE$ variants do not satisfy detailed balance. As a
result, we have no explicit $\mathcal{P}^{\ast }$s or effective $\mathcal{H}$%
s. Nevertheless, we use simple arguments concerning the likelihood of the
agents' actions and their effects on the collective behavior, often arriving
at good predictions for systems. Let us first specify the models and then
discuss their remarkable properties.

\textit{Egalitarian agents} (the $XIE_{egal}$ model) : Consider an extrovert
and its actions. Instead of randomly choosing an introvert (who is not
already connected to it) to add a link, it finds the \textit{least}
connected introvert to do so. This rule models an agent who realizes that
the introverts regard links as burden, and attempts to distribute this
burden as evenly as possible. Alternately, if we associate links with
wealth, then this agent's behavior can be thought of as giving wealth
(links) to the least fortunate. Thus, we coin the term `egalitarian' agent.
Similarly, an introvert would cut a link to the \textit{most} connected
extrovert, as this action would make the other extroverts more equal. Note
that PHS is still respected, since the rules in that language can be stated
simply as follows. An $I$ chooses the $E$ with maximal number of `particles'
to cut a link while an $E$ chooses an $I$ with the maximal number of `holes'
to add one. Finally, in case more than one partner satisfy the condition for
adding or cutting, then, our agent chooses one of those at random.

\textit{Elitist agents} (the $XIE_{elit}$ model) : Here, we consider the
opposite extreme. In this case, an extrovert prefers the most `sociable'
introvert, and adds a link to the \textit{most} connected of the available
introverts. In this sense, these agents award the wealthy, much like
groupies flocking to the most popular star. Similarly, an introvert cuts a
link to the \textit{least} connected available extrovert. Since all agents
keep the number of highly connected individuals (the elite) as large as
possible, we named them `elitists'. Again, PHS is respected, as we can
replace the word `maximal' above by `minimal' here.

Of course, we can study models where the introverts and extroverts are
selective in different ways, e.g., egalitarian $I$'s and elitist $E$'s. But,
to focus our investigation here, we will only consider the two cases above.
Since these has PHS, we need to run simulation for only system with, e.g., $%
N_{I}\leq N_{E}$, to compile data for both phases.

\begin{figure}[tbp]
\centering
\includegraphics[width=3.5in]{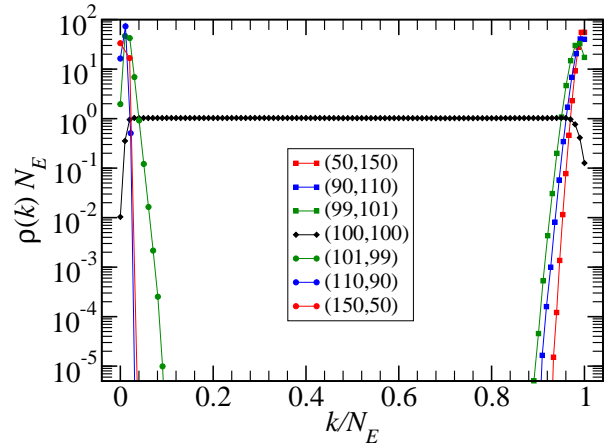}
\caption{Degree distribution of the introverts for various $XIE_{egal}$
systems with $200$ egalitarians: $\left( N_{I},N_{E}\right)
=(50,150),(90,110),(99,101),(100,100),(101,99),(110,90),$ and $(150,50)$. To
facilitate comparisons, we plot an appropriately scaled distribution, $%
\protect\rho N_{E}$, against $k/N_{E}\in \left[ 0,1\right] $. Apart from the
critical case, all distributions are confined to one or the other extreme.
}
\label{maxdd}
\end{figure}

\begin{figure}[tbp]
\centering
\includegraphics[width=3.5in]{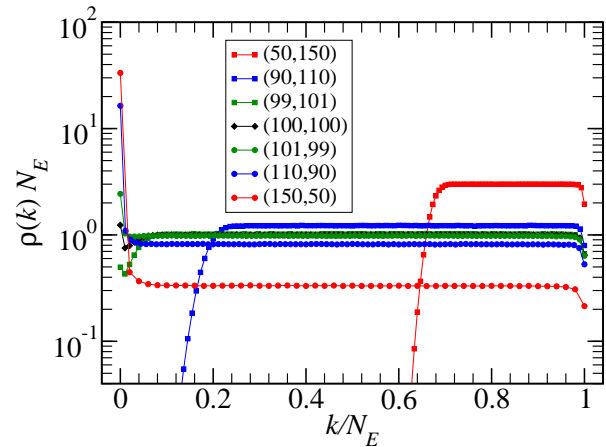}
\caption{Degree distribution of the introverts for various $XIE_{elit}$
systems with $200$ elitists: $\left( N_{I},N_{E}\right)
=(50,150),(90,110),(99,101),(100,100),(101,99),(110,90),$ and $(150,50)$. To
facilitate comparisons, we plot an appropriately scaled distribution, $%
\protect\rho N_{E}$, against $k/N_{E}\in \left[ 0,1\right] $. Note that the
plateau sections in the critical and the two nearby cases are too close to
be easily distinguished here.}
\label{mindd}
\end{figure}

Focusing on steady states, and to facilitate comparisons with previous data,
we study various systems with $N=200$ and $z=1$. All networks (apart from
some exceptions) have been initialized randomly and run for $10^{8}$ sweeps 
\footnote{%
With $\mathcal{N}$ attempts to update in a `sweep,' each link has an even
chance to change in a sweep.}. After discarding $10^{6}$ sweeps,
measurements on an agent of each community are taken after each sweep to
compile the degree distributions. Working with systems where $\left(
N_{I},N_{E}\right) =\left( 50,150\right) $, $\left( 90,110\right) $, $\left(
99,101\right) $, and $\left( 100,100\right) $, we measure degree
distributions of both the $I$'s and the $E$'s. From these, we extract
information relevant for both phases (using PHS). For convenience and
clarity, we show the data for only $\rho _{I}$ in a series that runs from $%
\left( 150,50\right) $ to $\left( 50,150\right) $, \textit{through} the
critical point. Plotted in Figs. \ref{maxdd} and \ref{mindd}, we see clearly
the effects of preferential attachment on the sharpness of the transition.
For $XIE_{egal}$, it is even sharper; for $XIE_{elit}$, the transition is
smooth. To form a more complete picture, we output typical configurations. 
\footnote{%
For example, see Figs. \ref{maxLong} and \ref{minLong} below.} These reveal
that, despite very similar (time averaged) degree distributions in critical
cases, the systems display drastically different behavior (at any given
time). Below we present the results, as well as our understanding, for each
variant.


\subsection{Steady state behavior for egalitarians}

In the low density regime of this $XIE_{egal}$ model, there are only a small
number of contacts. Since an extrovert has much bigger choice than an
introvert, the steady state properties are determined by the behavior of the 
$E$'s. Suppose we initialize the system with all links absent. As extroverts
add links and the fraction of introverts with a link rises, \textit{no} link
will be added to $I$'s with $k>0$, as long as there are some $I$'s with no
connections. The fraction of $I$'s with $k=1$ continues to increase till it
reaches a value ($\alpha =N_{E}/N_{I}$) when the rates for adding and
cutting are balanced. Thus, in the steady state, we have 
\begin{equation}
\rho _{I}(0)=1-\alpha ;~\rho _{I}(1)=\alpha .  \label{eq:egal}
\end{equation}%
As $\alpha $ increases, the fraction of isolated introverts decrease, so
that, when a rare fluctuation brings it to zero, the extroverts will create 
$I$'s with 2 links. In other words, this group of introverts will only occur
when extroverts are chosen much more often to update. We thus estimate that,
for fixed $\alpha <1$ and large $N$, $\rho _{I}(k\geq 2)$ decreases as 
$\mathcal{O}\left( e^{-b_{\alpha }N}\right) $, where 
$b_{\alpha }\rightarrow 0 $ as $\alpha \rightarrow 1$.

Turning to the extroverts, their degree distribution is less sharp and can
be obtained, again through another rate balance equation. For simplicity,
let us fix $z=1$. With degree $q$, there are $q/N$ chances for one of its $I$
contact to be chosen, leading to a drop in $q$. Since it will, when chosen,
definitely increase $q$, we are led to $\left( q/N\right) \rho _{E}\left(
q\right) =\left( 1/N\right) \rho _{E}\left( q-1\right) $ and the prediction 
\begin{equation}
\rho _{E}\left( q\right) =1/eq!
\end{equation}%
A similar argument can be made for $P\left( X\right) $. If $X\lesssim N_{I}$%
, introverts tend to have only one link or none. So, there are $X/N$ ways to
pick a connected $I$, who will cut its link. Yet, every choice of an $E$
leads to an increase in $X$, so that the rate balance equation reads $\left(
X/N\right) P\left( X\right) =$ $\left( N_{E}/N\right) P\left( X-1\right) $.
The result is a Poisson distribution $P\left( X\right) \propto N_{E}^{X}/X!$%
, exact in the $N\rightarrow \infty $ with fixed $\alpha $. For systems with
finite $\left( N_{I},N_{E}\right) $, it is possible for $X$ to exceed $N_{I}$
when $\alpha $ is close to unity. At the other extreme, for $X\gg N_{I}$,
every choice of $I$ will decrease $X$ . Thus, balancing the rates leads us
to a simple exponential, $P\left( X\right) \propto \alpha ^{X}$, in this
regime. A unified way to regard these regimes is that, as $X$ increases, the
ratio $P\left( X-1\right) /P\left( X\right) $ crosses over, from $X/N_{E}$
to $N_{I}/N_{E}$. The detailed nature of this cross-over may be quite
complex, but this general picture is borne out well in \textit{all} our data
(not shown). Finally, we may estimate the range of $\left(
N_{I},N_{E}\right) $ in which the Poisson distribution should hold. Since
the width of the Poisson distribution is of order $\sqrt{N_{E}}$, it should
be valid as long as $1-\alpha \lesssim {\mathcal{O}}\left( 1/\sqrt{N_{E}}%
\right) $. Thus, it is exact in the $N\rightarrow \infty $ with fixed $%
\alpha $. For the high density phase ($\alpha >1$), invoking PHS leads to
similar conclusions, namely, $\rho _{E}(N_{I})=1-\alpha ^{-1}$ and $\rho
_{E}(N_{I}-1)=\alpha ^{-1}$. Also, away from criticality, the variance in
the degree distribution is only ${\mathcal{O}}\left( 1\right) $. As a
result, the transition becomes much sharper than the proto $XIE$ model.
Finally, the critical case will be discussed extensively below, as it
exhibits the most interesting behavior.

\begin{figure}[tbp]
\centering
\includegraphics[width=3.5in]{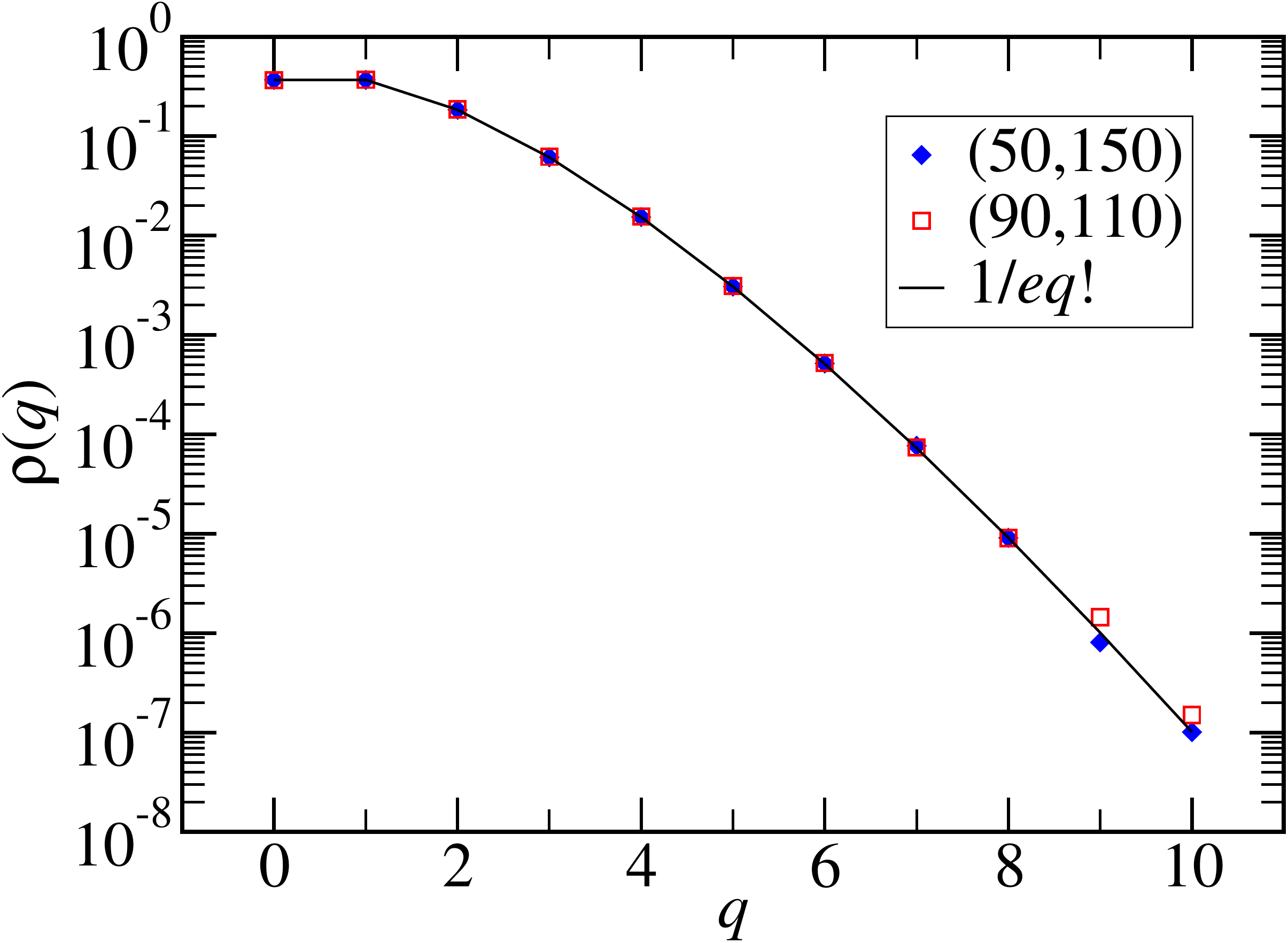}
\caption{Degree distributions for egalitarian extroverts, $\protect\rho %
_{E}\left( q\right) $, far in the low density phase.}
\label{maxDists-a}
\end{figure}

From Fig. \ref{maxdd}, we see various aspects of the expected, the most
prominent being distributions confined to one of the two extremes when 
$N_{I}\neq N_{E}$. At the more detailed level, the predictions 
(\ref{eq:egal}) is in perfect agreement with the 
$\left( 150,50\right) $ data: 
$\protect\rho _{I}\left( 0\right) =0.6667025$ and 
$\protect\rho _{I}\left(1\right) =0.3332975$.
Meanwhile, though $\rho_{I}\left( 0\right) =1-\alpha $ is very well 
satisfied for \textit{all} low density cases, the data shows a
detectable spread beyond $k=1$ (except for the 
$\left( 150,50\right) $ case). For the near critical system, $\left(
101,99\right) $, two features in $\rho _{I}$ are noteworthy. For $k\geq 2$,
it drops exponentially. With an extremely well fit to $e^{-2k}$, it behooves
us to conjecture $\rho _{I}\propto e^{-\Delta k}$. Before this decay, $\rho
_{I}$ rises substantially, from $0.198$ ($\cong 2/101$) to $0.473$ and $%
0.427 $ for $k=1$ and $2$, respectively. A good explanation for this
behavior is yet to be advanced. By contrast, the distributions for the
extroverts, $\rho _{E}\left( q\right) $, in this regime (related, through
PHS, to $\rho _{I}$ in the high density phase shown in Fig \ref{maxdd}) fit
well to the prediction above, $1/eq!$, as shown in Fig \ref{maxDists-a}.


\begin{figure}[tbp]
\centering
\includegraphics[width=3.5in]{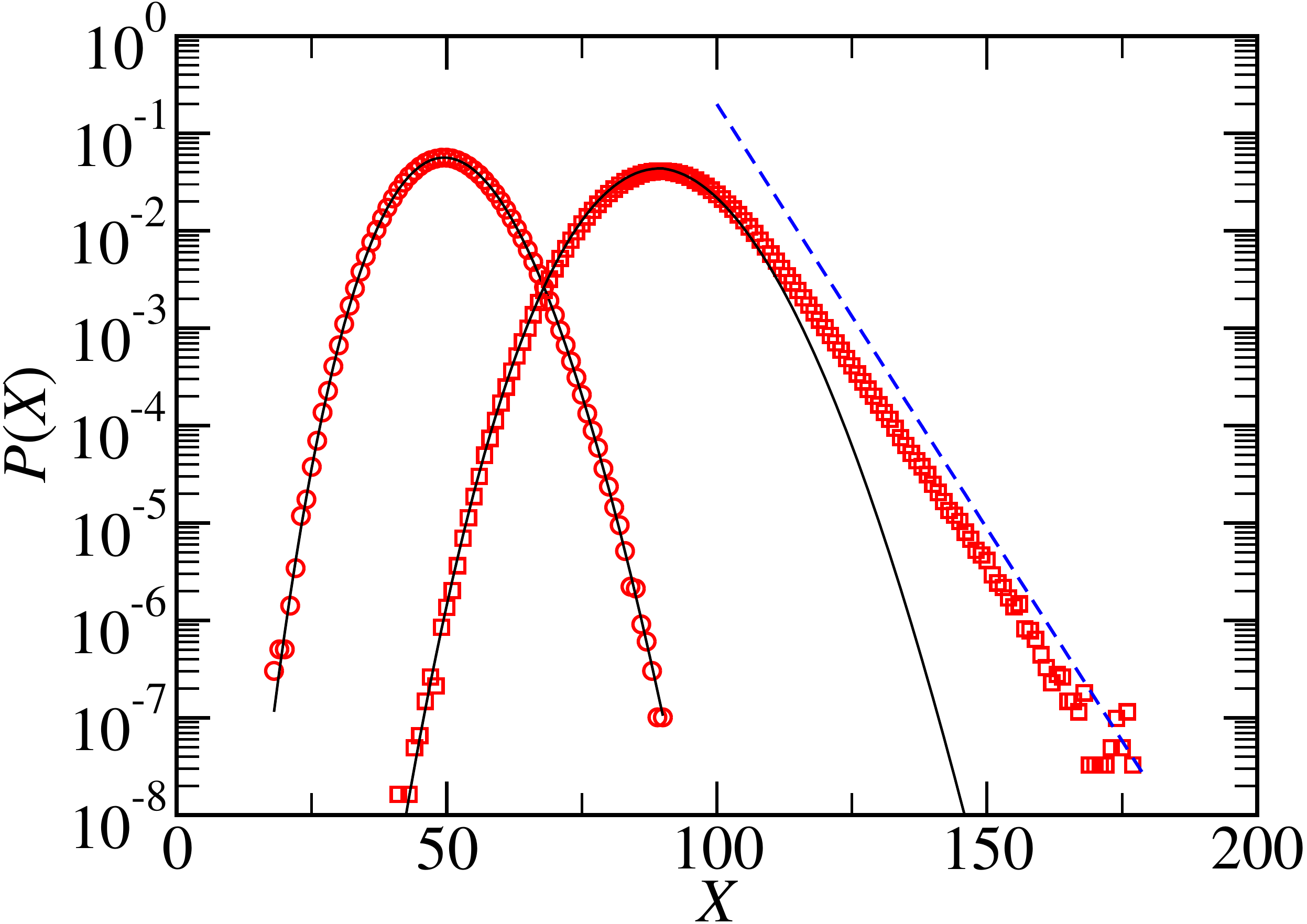}
\caption{Distributions for $X$ in two low density $XIE_{egal}$ systems: 
$\left( 150,50\right) $ (red circles) and $\left( 110,90\right) $ 
(red squares).
Appropriate Poisson distributions are shown as black lines. The dashed blue
line is the exponential, $\left( 9 / 11 \right)^X $, provided as a guide to
the eye.}
\label{maxDists-b}
\end{figure}


Let us turn to the data for the distributions $P\left( X\right) $ away from
criticality. Fig \ref{maxDists-b} shows that our prediction agrees with the $%
\left( 150,50\right) $ case perfectly. It is clear that the next case $%
\left( 110,90\right) $ is displaying cross-over behavior, as the Poisson
distribution provides a reasonable fit only for $X\lesssim 110=N_{I}$.
Beyond that, $P$ decays asymptotically into an exponential. To guide the
eye, we plotted $\alpha ^{X}$ as a dashed line (with $\alpha =90/110$ here).
The validity of this cross-over scenario persists (not shown here) to the
near critical case $\left( 101,99\right) $, in which $\left( 99/101\right)
^{X}$ provides an excellent fit in a wide range of $X$ (from $\thicksim 150$
to the maximum accessible by our computers, $\thicksim 900$). To summarize
the off-critical data in this variant, we conclude that the extreme Thouless
effect is even stronger, as $m\left( h,N\right) -sign\left( h\right) $
appears to vanish as $e^{-N}$ for any $h\neq 0$.

The critical case ($N_{I}=N_{E}\equiv L$) is the most fascinating. In Fig. %
\ref{maxdd}, we see that $\rho _{I}$ is statistically flat across the entire
range, except for dips at both ends. This behavior might be expected from
the phrase we coined, `egalitarian,' but such a completely flat distribution
does not reveal the typical configurations as the system evolves. In
particular, under the egalitarian dynamics, an agent with degree $k$ will
not get an added link, if there is just a single agent with degree $k-1$.
Thus, at any particular time $t$, most agents are expected to have degrees
in a narrow range, say $\pm 1$, around some value, ${\mathcal{\lambda }}%
\left( t\right) $. In this sense, the egalitarian agents, acting together,
do achieve their aim of minimizing within population variations. However,
since there is no bias in favor of either group, we expect that ${\mathcal{%
\lambda }}\left( t\right) $ should perform an unbiased random walk, within $%
\left[ 0,L^{2}\right] $, over long times. This phenomenon is confirmed by
simulations, as discussed next.

\begin{figure}[tbp]
\centering
\includegraphics[width=3.5in]{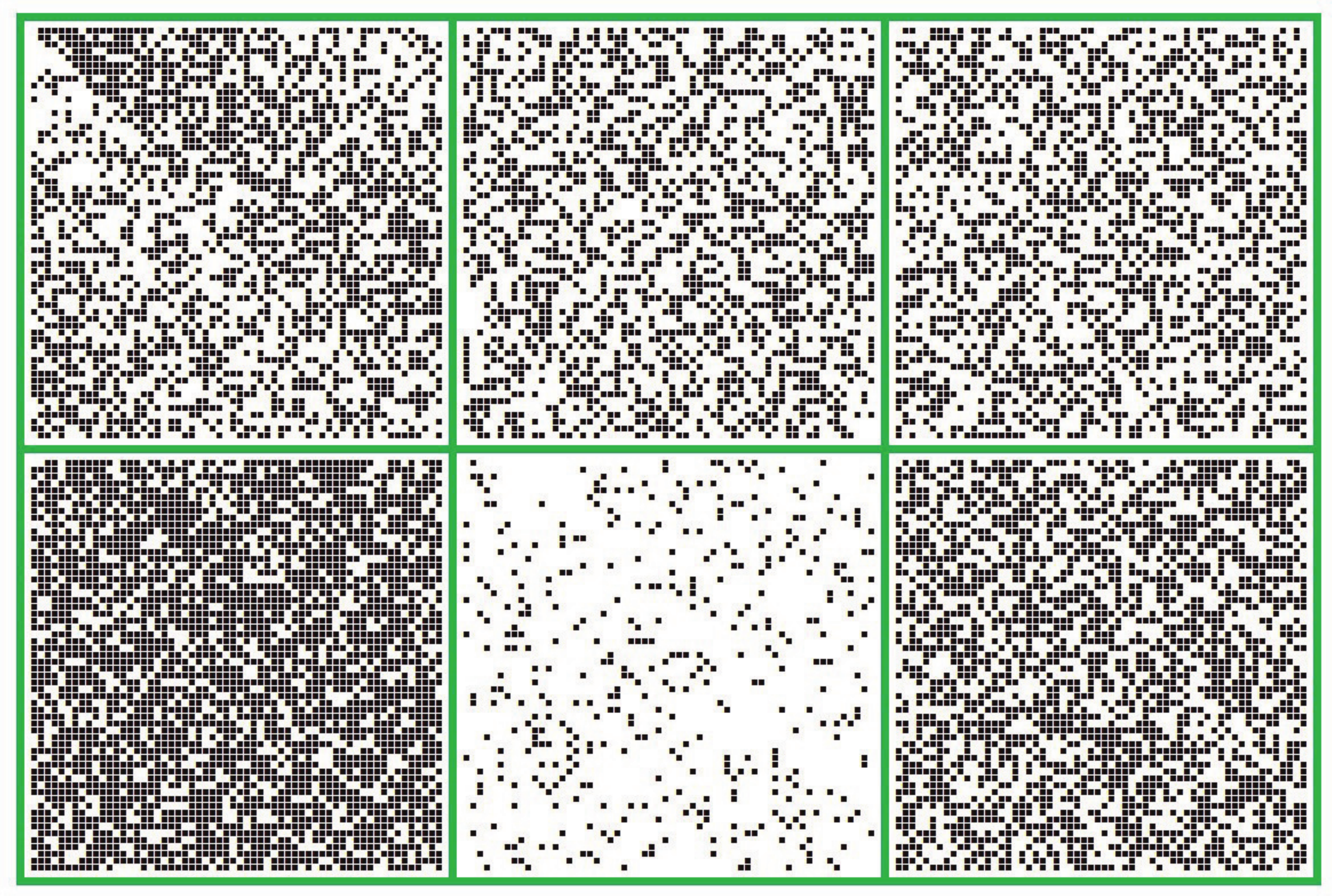}
\caption{ Evolution of typical incidence matrices for a $\left( 60,60\right) 
$ system of egalitarians. The initial configuration is fully ordered, with
links above the diagonal. A snapshot is taken after $10^{s}$ sweeps. The
panels here, from the left, are snapshots at $s=0,1,\ldots ,5$. In each
snapshot shown, the degree of every agent is essentially the same: $%
30,27,26,44,7,$ and $34$, respectively.}
\label{maxLong}
\end{figure}

\begin{figure}[tbp]
\centering
\includegraphics[width=3.5in]{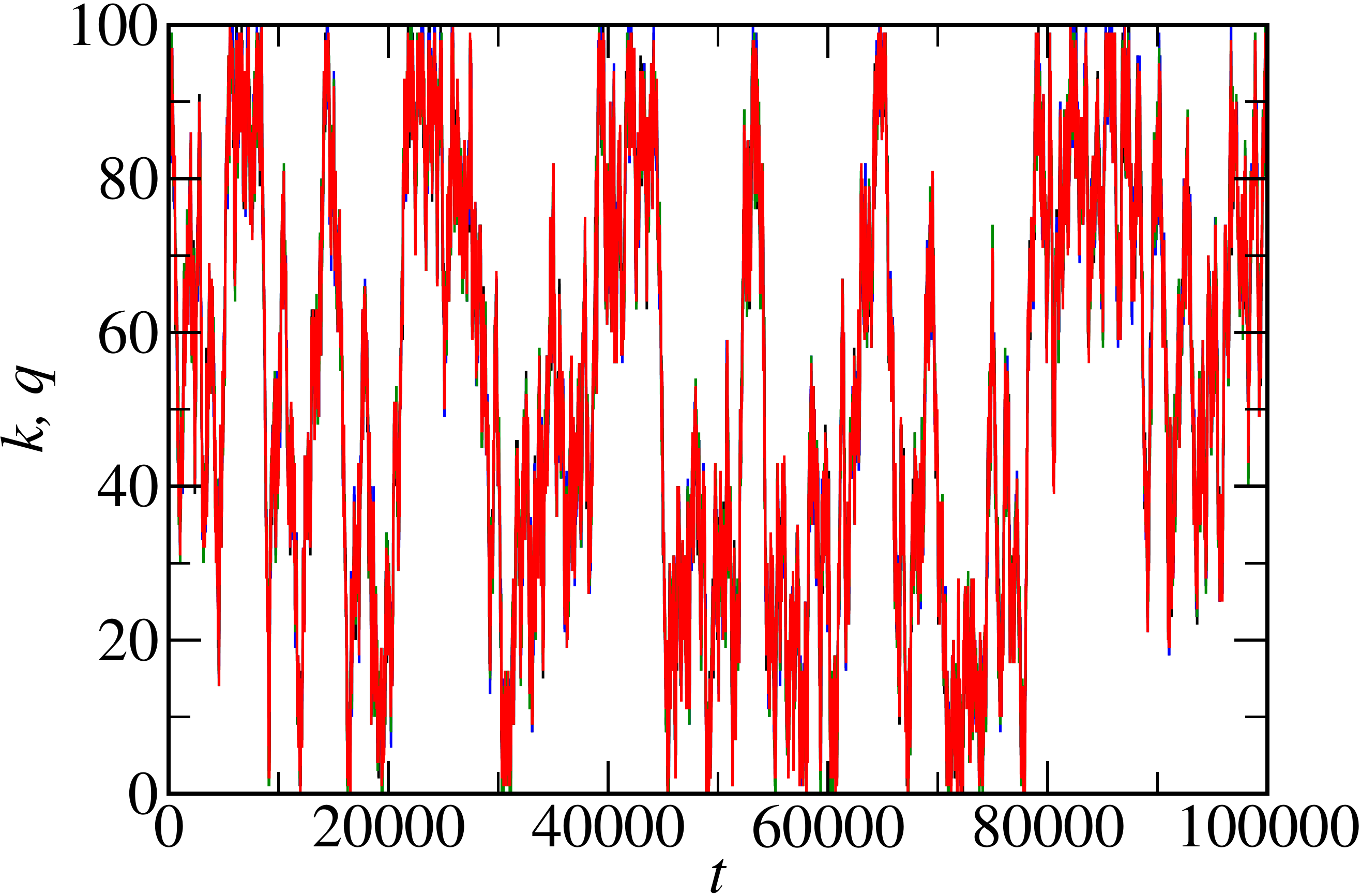}
\caption{Time trace of the degrees of two introvert egalitarians, $k$
(black, blue), and of two extroverts, $q$ (red, green) in a critical $\left(
100,100 \right) $ system. Here the unit of $t$ is a sweep. Note the typical
time for traversing the full range is $\mathcal{O}\left( L^{2}\right) $
sweeps. It is difficult to resolve the 4 traces, even when a small portion
is magnified (Fig. \protect\ref{4maxBIG}).}
\label{4maxagents}
\end{figure}

To pre-empt possible critical slowing down, we carry out very long runs ($%
5\times 10^{8}$ sweeps) with a smaller system: $\left( 60,60\right) $.
Initializing the incidence matrix to be a fully occupied upper triangle (not
significant here, but more so below), we output $\mathbb{N}$'s at $10^{s}$ ($%
s=0,...,5$) sweeps. These are shown in Fig. \ref{maxLong} (from the top
left; a black/white square represents a present/absent link). While every
sample appears disordered, note how the total $X$ differs considerably from
one $s$ to the next. In Fig. \ref{4maxagents}, we show the time traces of
the degrees of two introverts and two extroverts (after the system is
relaxed $2\times 10^{s}$ sweeps). The four traces are almost
indistinguishable, as expected. Over time, they perform the same random
walk, over the entire available range. To quantify such behavior, we may
define the average degree at any instant $t$, by%
\begin{equation}
\lambda \left( t\right) \equiv X\left( t\right) /L
\end{equation}
where $X$ is of course the same as $\Sigma _{i}k_{i}$ of the introverts and $%
\Sigma _{j}q_{j}$ of the extroverts, as well as and two variances%
\begin{equation}
D_{I}\left( t\right) \equiv \frac{1}{L}\sum_{i}k_{i}^{2}-\lambda
^{2};~~D_{E}\left( t\right) \equiv \frac{1}{L}\sum_{j}q_{j}^{2}-\lambda ^{2}
\end{equation}%
On the one hand, we expect $D_{I,E}(t)$ to vary little in time and remain $%
\mathcal{O}\left( 1\right) $ (even at this critical point). Thus, the time
averaged $D$'s will also be $\mathcal{O}\left( 1\right) $. In stark
contrast, $\lambda \left( t\right) $ performs the same random walk as $%
X\left( t\right) $. Thus, the time averaged distribution is flat over most
of the full range $\left[ 0,L\right] $ (in Fig. \ref{maxdd}), leading to a
variance of $\mathcal{O}\left( L^{2}\right) $. Far from the extremes, $X$
changes by $\pm 1$ at each attempt with equal probability and so, the time
scale for traversing $L^{2}$ is $\mathcal{O}\left( L^{2}\right) $ sweeps ($%
\mathcal{O}\left( L^{4}\right) $ attempts). This estimate is entirely
consistent with the traces in Fig. \ref{4maxagents}. To summarize, a simple
picture emerges for the critical dynamics. At short time scales, the
egalitarian practice of agents ensures a sharply peaked distribution, with
each agent having only one link more than, or less than, some value 
$\lambda$. Over longer periods, $\lambda $ wanders, 
as $X/L$ arrives close to an
integer. Further studies should provide a more detailed and quantitative
picture of the remarkable collective behavior of such a minimal model.


\ 

\begin{figure}[tbp]
\centering
\includegraphics[width=3.5in]{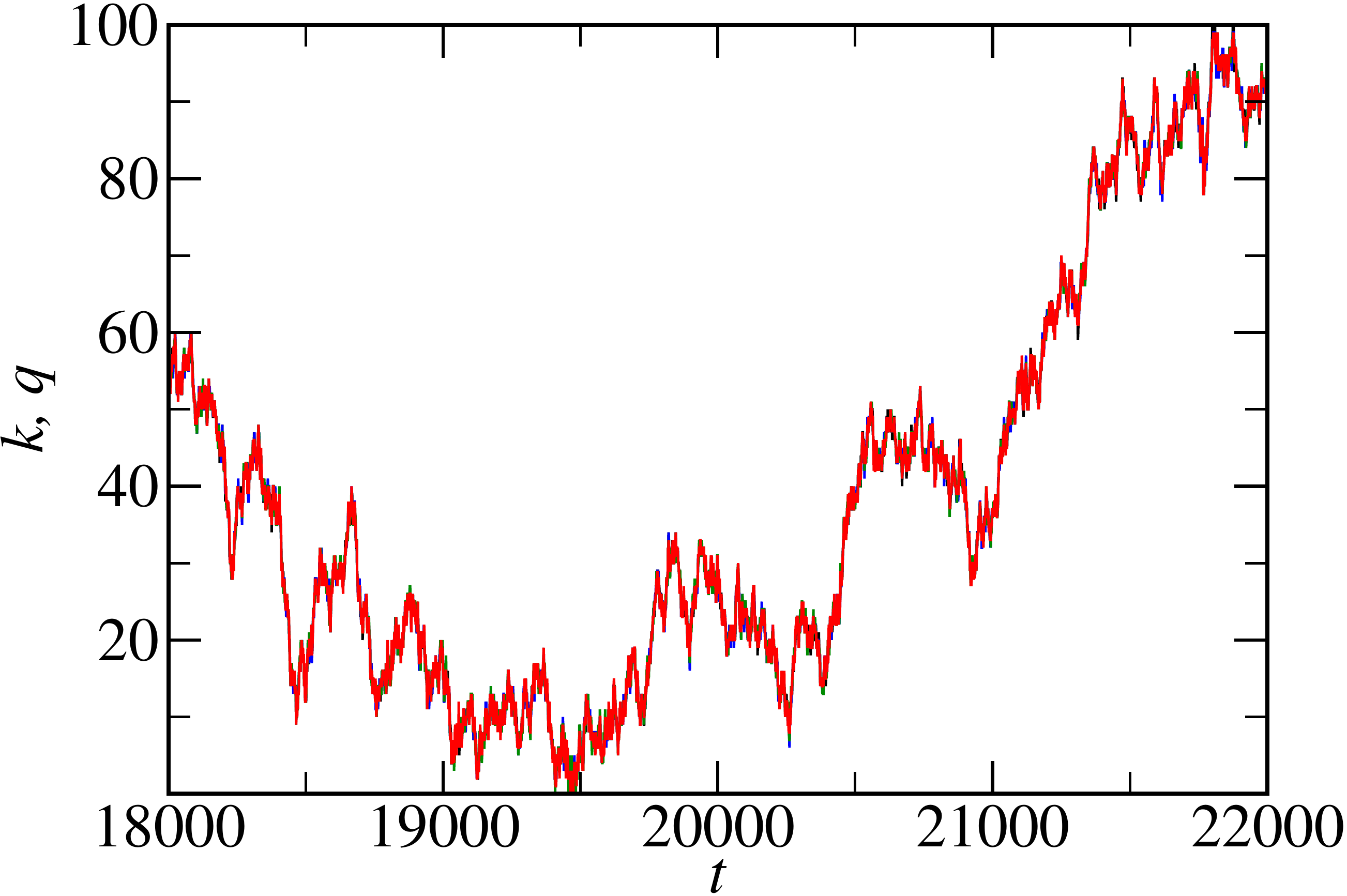}
\caption{An expanded region of Fig. \protect\ref{4maxagents}. Symbols are
the same here.}
\label{4maxBIG}
\end{figure}

\subsection{Steady state properties for elitists}

Turning to the $XIE_{elit}$ model, we again consider the low density phase
first. Starting from a random distribution of introvert degrees, an
interesting instability should set in. The introvert with the largest degree
will be selected for attachment as soon as \textit{any} extrovert (not
already linked to it) is chosen to act. Since its $k$ can decrease only when
it is selected, its degree is likely to rise rapidly. Of course, $k$ cannot
rise beyond $N_{E}$, and the extroverts will pick the next `star' for
attachment. This instability will continue until the steady state, in which
the rate of links being added equals the rate for deletion. In a low density
phase, the probability that an extrovert is fully connected vanishes for
large $N$, so that it can always add a link if selected. Thus, the rate for
adding links is just $N_{E}/N$. But the rate for deletion is proportional to
the fraction of introverts with one or more links. Balancing the above rate
with this, $\left[ 1-\rho _{I}\left( 0\right) \right] N_{I}/N$, we find the
fraction of isolated introverts to be $1-\alpha $, an exact result for large 
$N$. Making a similar argument for the high density phase, we expect a
smooth cross-over, with no discontinuities, as $\alpha $ is varied through
unity. Simulations confirm this picture. Yet, the degree distributions
display unexpected unique properties, as illustrated in Fig. \ref{mindd}.
For the introverts, we see that the data in the first two cases for 
$\protect\rho _{I}\left(0\right)$, $0.667436$ and $0.182119$, agree
with the predicted value $1-\alpha$ to 0.2\% . Meanwhile, the rest
of the introverts appears to be connected in an unusual way, with $\rho
_{I}\left( k\right) $ distributed evenly, up to almost the maximum $N_{E}$.
At first sight, a flat $\rho $ reminds us of the critical egalitarian case.
However, such a plateau can be realised in another manner. In the next
paragraphs, we will show a remarkable underlying structure, not easily
discernible by examining the incidence matrix, $\mathbb{N}$.

\begin{figure}[tbp]
\centering
\includegraphics[width=3.5in]{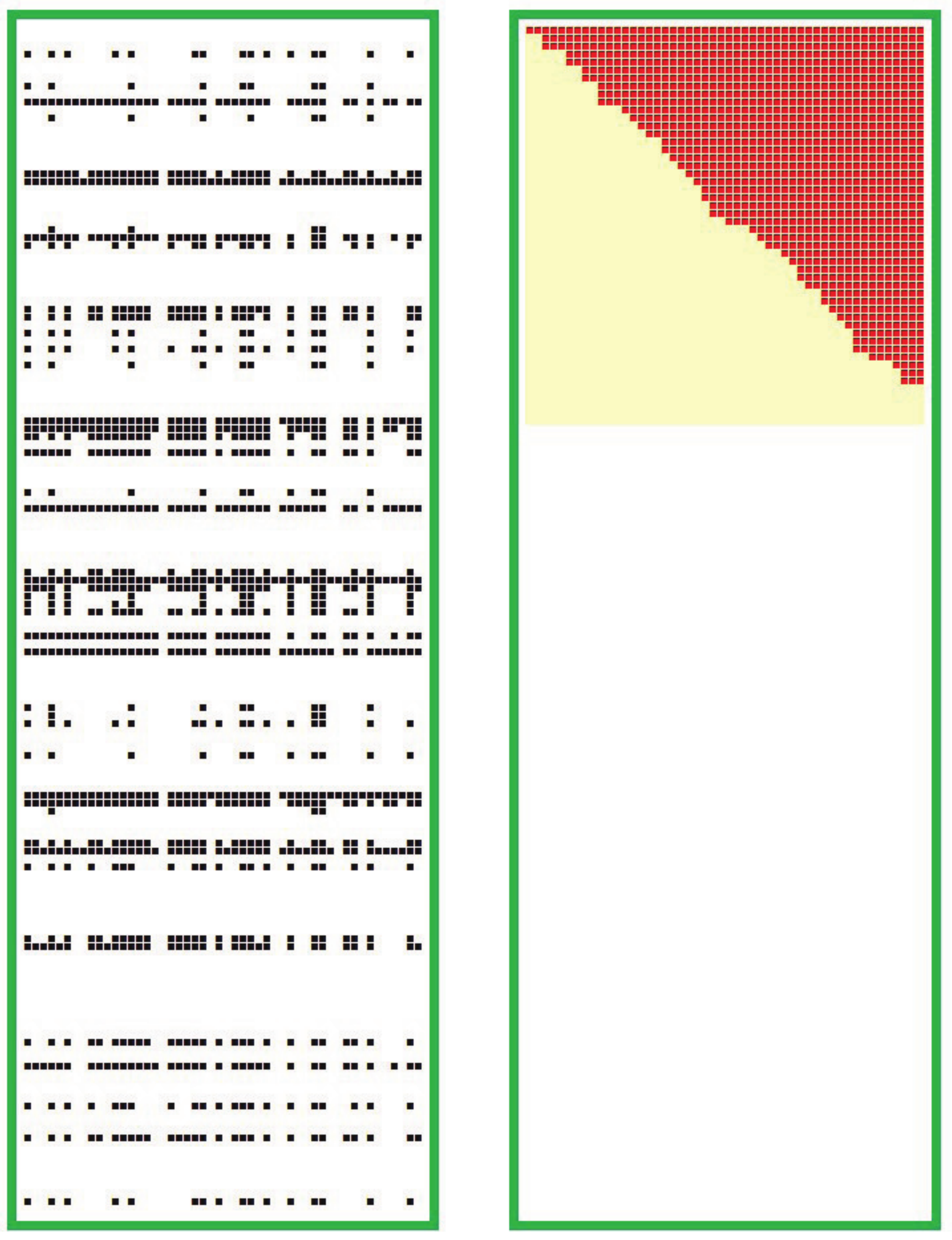}
\caption{Typical snapshot of an incidence matrix for a $\left(
N_{I},N_{E}\right) =\left( 150,50\right) $ $XIE_{elit}$ system in the steady
state (left). The sorted $\mathbb{N}$ is shown in red (right). The yellow
region is a $50\times 50$ square, as a guide to the eye. The green border
serves to indicate the extent of the matrix.}
\label{min15050}
\end{figure}

\begin{figure}[tbp]
\centering
\includegraphics[width=3.5in]{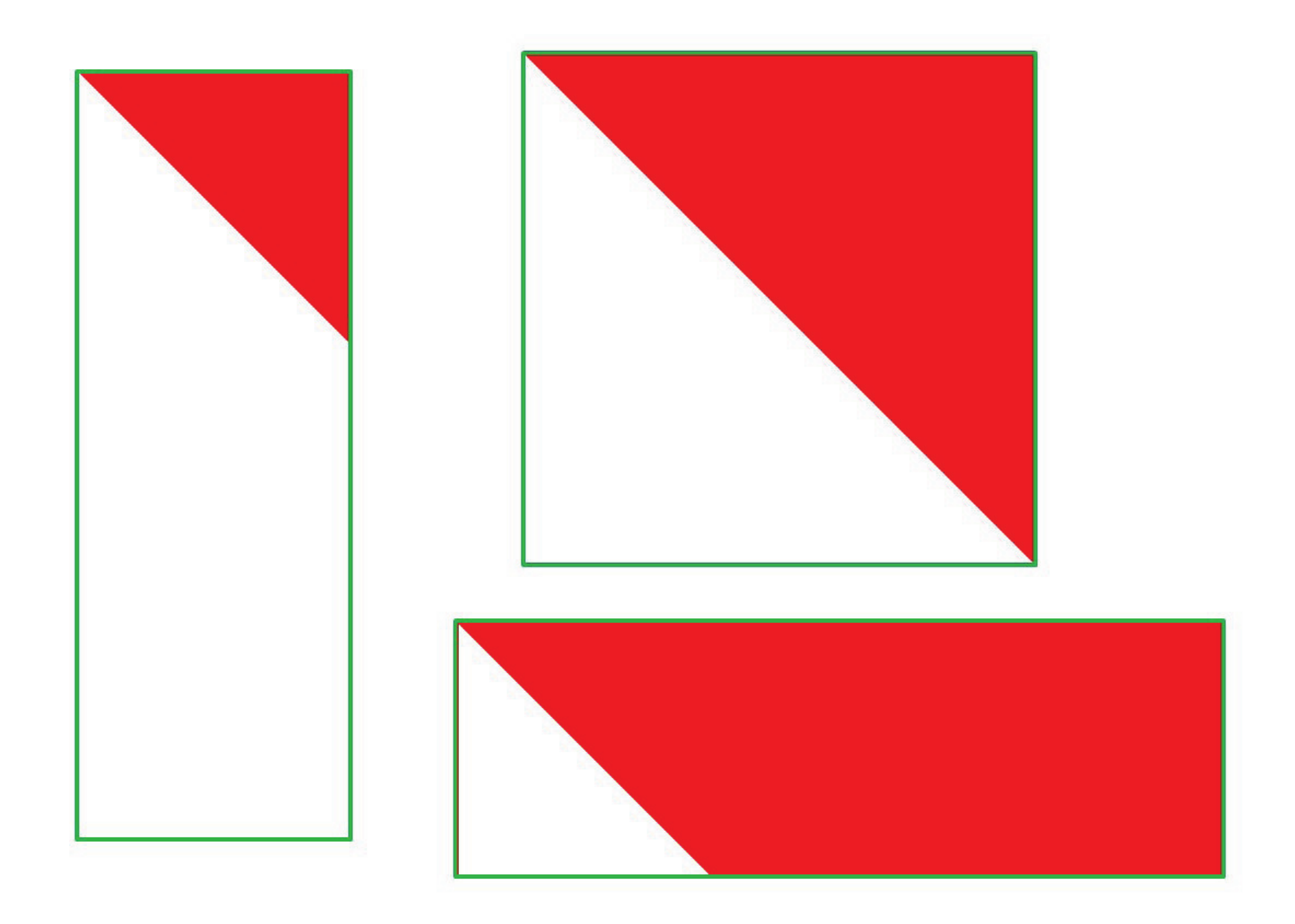}
\caption{ Schematic of typical steady state configurations (incidence
matrices) in the $XIE_{elit}$ model -- after sorting. In all cases, there
are $N_{I}$ rows and $N_{E}$ columns. The red (white) region denotes the
presence (absence) of a link. The rectangle on the left (right) represents
the case of $N_{I}=150$ ($50$) and $N_{E}=50$ ($150$). The square represents
the critical system: $N_{I}=N_{E}=100$. }
\label{minall}
\end{figure}

Consider a typical configuration for the $\left( 150,50\right) $ case,
illustrated by $\mathbb{N}$ on the left in Fig. \ref{min15050} (again, black
and white entries represent present and absent links, respectively). The
links appear scattered throughout the matrix. However, considerable order is
revealed if we `sort' the agents as follows. On the right of Fig. \ref%
{min15050}, we show the same $\mathbb{N}$, with the rows and columns
permuted into an ordered list, in which the most connected introvert
(extrovert) is placed on the top row (right column). To avoid confusion, we
use red (squares for a link) for such `sorted' $\mathbb{N}$'s. While indeed, 
$2/3$ (i.e., $1-N_{E}/N_{I}$) of the rows are empty, the remaining connected
agents arrange themselves in an orderly fashion. To guide the eye, we shaded
this region ($50\times 50$) yellow and roughly, an upper triangular matrix
emerges here. Clearly, a strictly ordered $L\times L$ matrix of this type
will produce a completely flat distribution: $\rho \left( k\right) \thicksim
1/L$. The insight gained here allows us to interpret the rest of the
distributions in Fig. \ref{mindd}. When the agents are sorted, the $\mathbb{N%
}$'s will progress, as schematically sketched in Fig. \ref{minall}, from the 
$\left( 150,50\right) $ rectangle on the left, through the $\left(
100,100\right) $ square, to the $\left( 50,150\right) $ rectangle on the
right. The best summary of such behavior is: All agents in the minority will
partner with a similar number of those in the majority, creating a
triangular incidence matrix (after sorting), while the rest of the majority
are static and content (with all links or none). Of course, each individual
changes partners often, but the sorted network displays little variation.
Given this picture, it is easy to see that $\left\langle X\right\rangle $
varies continuously through criticality. Indeed, we can easily predict the
fraction $\left\langle x\right\rangle $, using this area of red region ($%
\left\vert N_{E}-N_{I}\right\vert +\left( N_{E}-N_{I}\right) ^{2}/2$), and
arrive at an equation of state: 
\begin{equation}
m\left( h\right) =\frac{2h}{1+\left\vert h\right\vert }
\end{equation}%
While the singularity is undoubtedly smoothed out in finite systems, this
result is most likely exact in the $N\rightarrow \infty $ limit. This
formula certainly captured the essence of the model: It predicts $0.0198$, $%
0.1818$, $0.6667$ for the three non-critical low density systems here, while
the observed values are $0.0207$, $0.1819$, and $0.6662$, respectively.
Though the plateau and triangular $\mathbb{N}$'s are the most prominent
features, there are other noteworthy details in Fig. \ref{mindd} : Both the
dips at the ends of each plateau and the non-monotonic behavior in $\rho
_{I}\left( k\thicksim 0\right) $ for the $N_{I}\gtrsim N_{E}$ systems are
intriguing. Clearly, there is ample room for theoretical explanations.


\begin{figure}[tbp]
\centering
\includegraphics[width=3.5in]{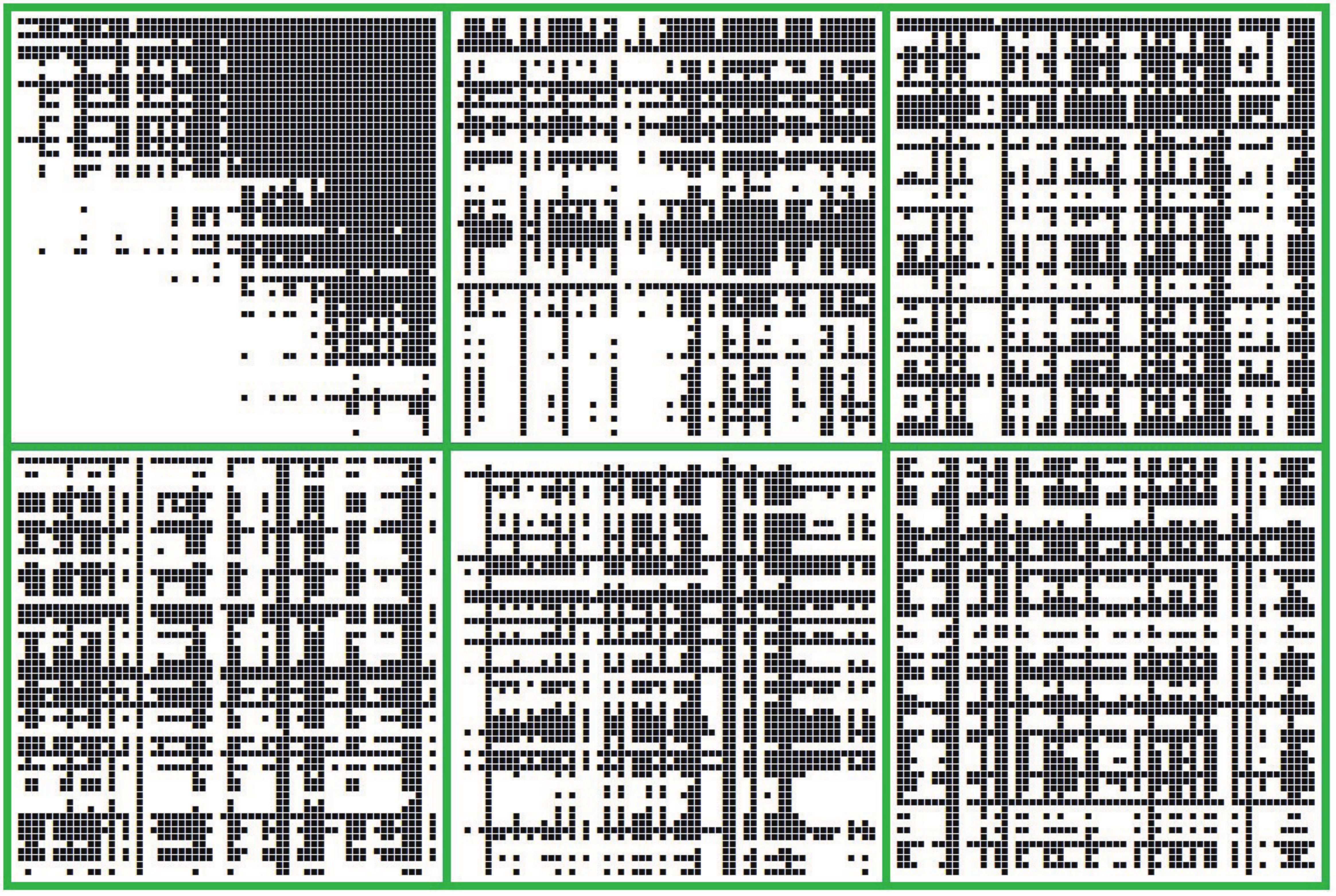}
\caption{Evolution of typical incidence matrices for a $\left( 60,60\right)$
system of elitists (the $XIE_{elit}$ model). The initial configuration is
fully ordered, with links above the diagonal. A snapshot is taken after $%
10^{s}$ sweeps. The panels here, from top left, are snapshots at $%
s=0,1,\ldots ,5$. When sorted, all of them resemble closely the initial
configuration. See Fig. \protect\ref{minmax10^5} for an example ($s=5$) of
the sorted matrix.}
\label{minLong}
\end{figure}

\begin{figure}[tbp]
\centering
\includegraphics[width=3.5in]{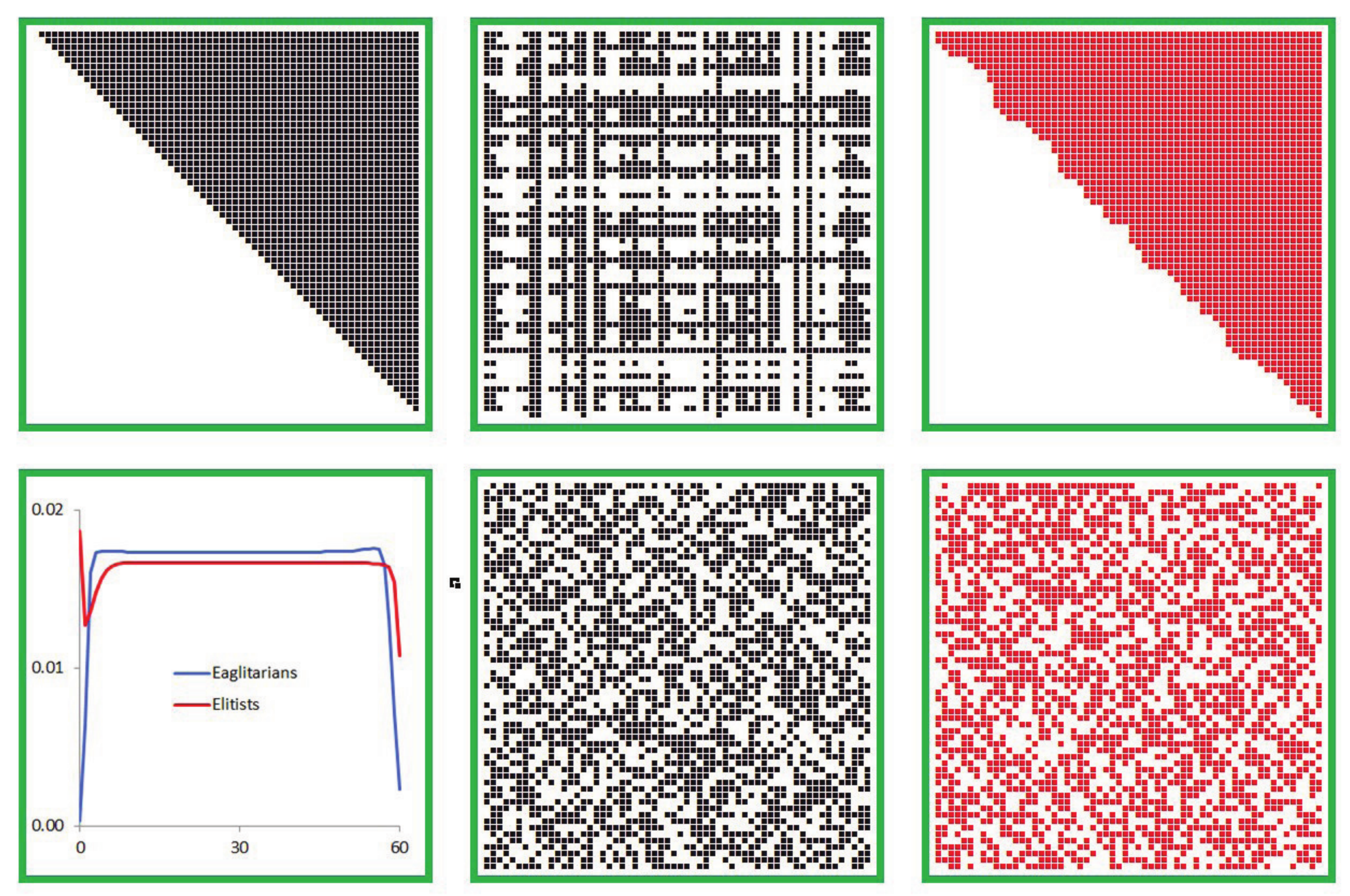}
\caption{Incidence matrices, initially (top left) and after $10^{5}$ sweeps
(center). The latter, when sorted, are shown in red (right). Top/bottom row:
elitists/egalitarians. Both stationary degree distributions are shown in the
lower left panel.}
\label{minmax10^5}
\end{figure}

Though the degree distributions are not expected to show `large
fluctuations,' the typical $\mathbb{N}$s (e.g., left panel in 
Fig. \ref{min15050} ) raise a
different question, concerning the nature of disorder. For simplicity, let
us explore the time evolution of the critical case, where all but a small
fraction of agents should be `active'. Expecting the configuration shown in
the schematic sketch, we start with an ordered state ($n_{i<j}=1$ and $0$
otherwise; fully occupied upper triangle in a $60\times 60$ square). As
above, we output an $\mathbb{N}$ after $10^{s}$ sweeps, for $s=0,...,5$. In
Fig \ref{minLong} we see that, even after a single sweep (upper left panel),
significant `disorder' already appears. Though disorder seems to increase
steadily, we find that the sorted $\mathbb{N}$ remains approximately the
same throughout the evolution. As an example. we show in Fig \ref{minmax10^5}
the configuration at the last time (top central panel) along with its sorted
version (top right panel). Indeed, it is straightforward to show that, if we
resort the agents after each update, it is not possible for an introvert to
create a $0$ (cut a link) in the sea of $1$s, or for an extrovert to create
a $1$ in the domain of $0$s. Thus, the staircase like interface between the
two domains will be preserved at each step and our $XIE_{elit}$ can be
mapped into a 1-dimensional interface, starting at the top left corner of
the square and ending at the lower right. The configurations are readily
labeled by a string of $L$ vertical and $L$ horizontal steps, e.g., 
$VHHVVVVHVH...$. With this mapping, we see that the rules of evolution are
simple: Choose a step (an element of the string) at random and exchange the
first unlike pair to its right. For example, if the third $V$ is chosen, the
new string is $VHHVVVHVVH...$. We see that the unlike pairs are exchanged
with varying rates that depend on the length of the domain to the left.
This mapping also reduces the number of configurations considerably, from 
$2^{L^{2}}$ for all possible $\mathbb{N}$s to just $\binom{2L}{L}$. 

Turning to non-critical cases, we see that the relevant strings consist of 
$N_{I}$ $V$s and $N_{E}$ $H$s, so that we have a total of 
$\binom{N_{I}+N_{E}}{N_{I}}$ configurations. 
As a result of the dynamics, there is a high
probability that the strings end with a domain of the majority, of length 
$\thicksim \left\vert N_{I}-N_{E}\right\vert $. 
For example, in the $\left(150,50\right) $ case (Fig. \ref{min15050}), 
we will find that most of the
activity takes place within the first $50+50$ elements of the string,
leaving $\thicksim 100$ $V$s essentially static. Of course, the active part
of the string here corresponds to the yellow region in the right panel of 
Fig. \ref{min15050}. 

Despite the significant reduction of configurations (from $2^{N_{I}N_{E}}$
to $\binom{N_{I}+N_{E}}{N_{I}}$) and simplification of the rules, the
dynamics for this interface model does not satisfy detailed balance, so that
finding the stationary distribution of this interface will be challenging.
Nevertheless, by considering equivalent classes of $\mathbb{N}$s (from
sorting) in the $XIE_{elit}$ model, this mapping should be a promising
approach for a better understanding of the behavior of our network of
elitist introverts and extroverts.


To end this section, let us highlight the dramatically different behaviors
between the two variants when $N_{I}=N_{E}$. In the central panels Fig.\ref%
{minmax10^5}, we show the last configurations (after $10^{5}$ sweeps) of
both systems, starting from the same initial $\mathbb{N}$ (top left panel).
After sorting, these matrices take the widely disparate forms shown in the
right panels (in red). Meanwhile the time-averaged degree distributions of
both are, apart from minor differences at the two extremes, practically 
\textit{identical} (lower left panel)! The time traces of four specific
agents also reveals the major differences. Illustrated in Fig. \ref{4maxBIG}
is a small portion of Fig. \ref{4maxagents}, where we see that indeed, the
egalitarian agents have essentially the same degree, $\lambda $, at any
time, but that $\lambda $ wanders over the full range. By contrast, Fig. \ref%
{4minagents} and a magnified portion, Fig. \ref{4minBIG}, clearly show that
these elitist agents have wildly differing degrees in general, but each
agent's degree wanders over the full range. In the society of `elitists,'
the inequality at any time is quite extreme. Yet, over time, any particular
individual experiences both extreme `affluence' and `poverty.' The time
averaged `wealth' of all individuals is the same, and in this sense,
symmetry is restored as all agents must be equal on the average. Similar to
the egalitarian case, there is little variation in a different, more subtle,
aspect here. When ranked by the number of connections, these individuals
fall into the same order. In other words, at any time, a specific
permutation ($\pi $) of the agents will expose an ordered state, but it is $%
\pi $ that wanders over long periods. There are clearly substantial
correlations between \textit{all} agents in both cases. These fascinating
aspects should provide much food for thought and many avenues for future
explorations. 

\begin{figure}[tbp]
\centering
\includegraphics[width=3.5in]{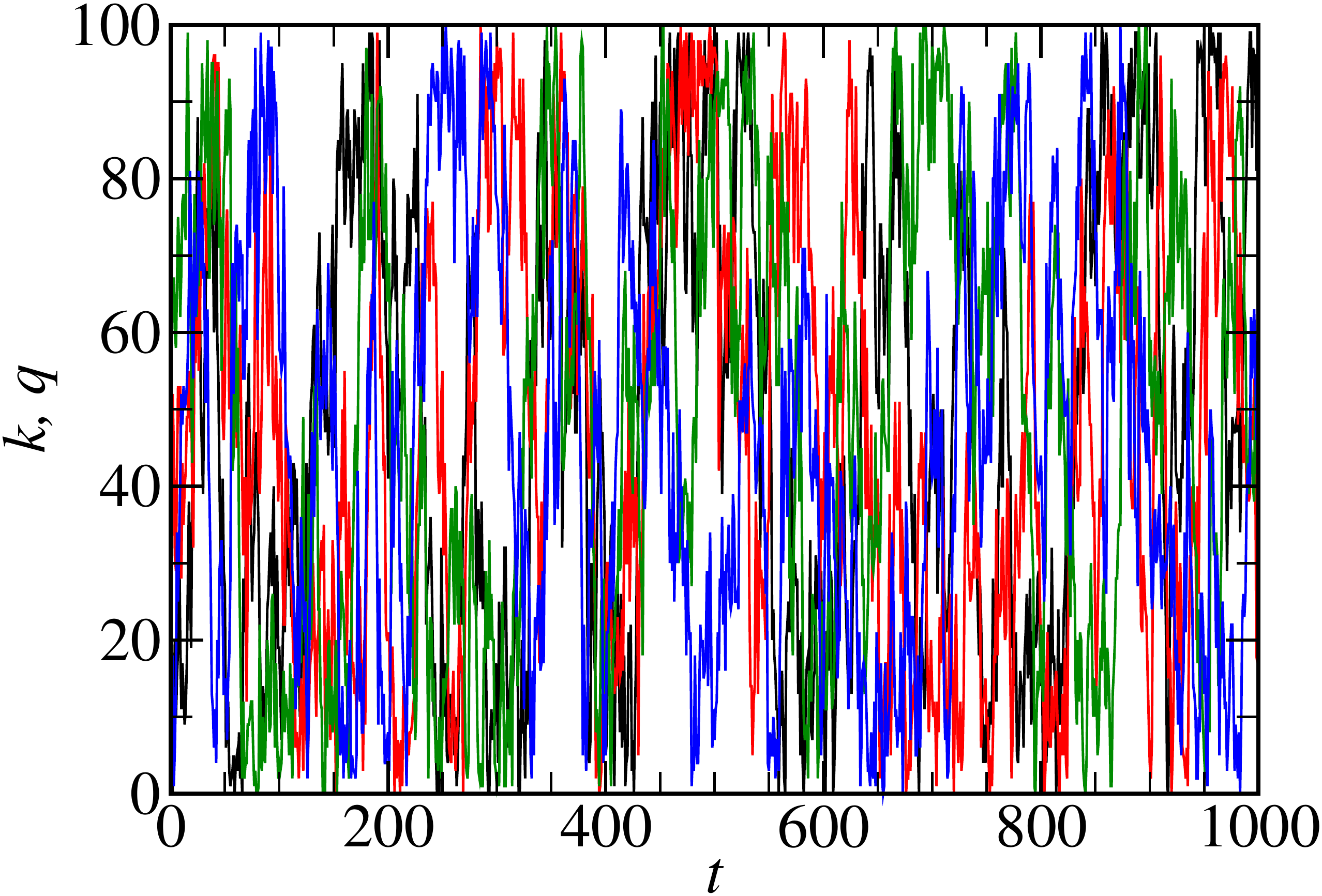}
\caption{Time trace of the degrees of two introvert elitists, $k$ (black,
blue), and of two extroverts, $q$ (red, green) in a critical $\left( 100,100
\right) $ system. Note the four traces are very different, unlike the
egalitarians case.}
\label{4minagents}
\end{figure}

\begin{figure}[tbp]
\centering
\includegraphics[width=3.5in]{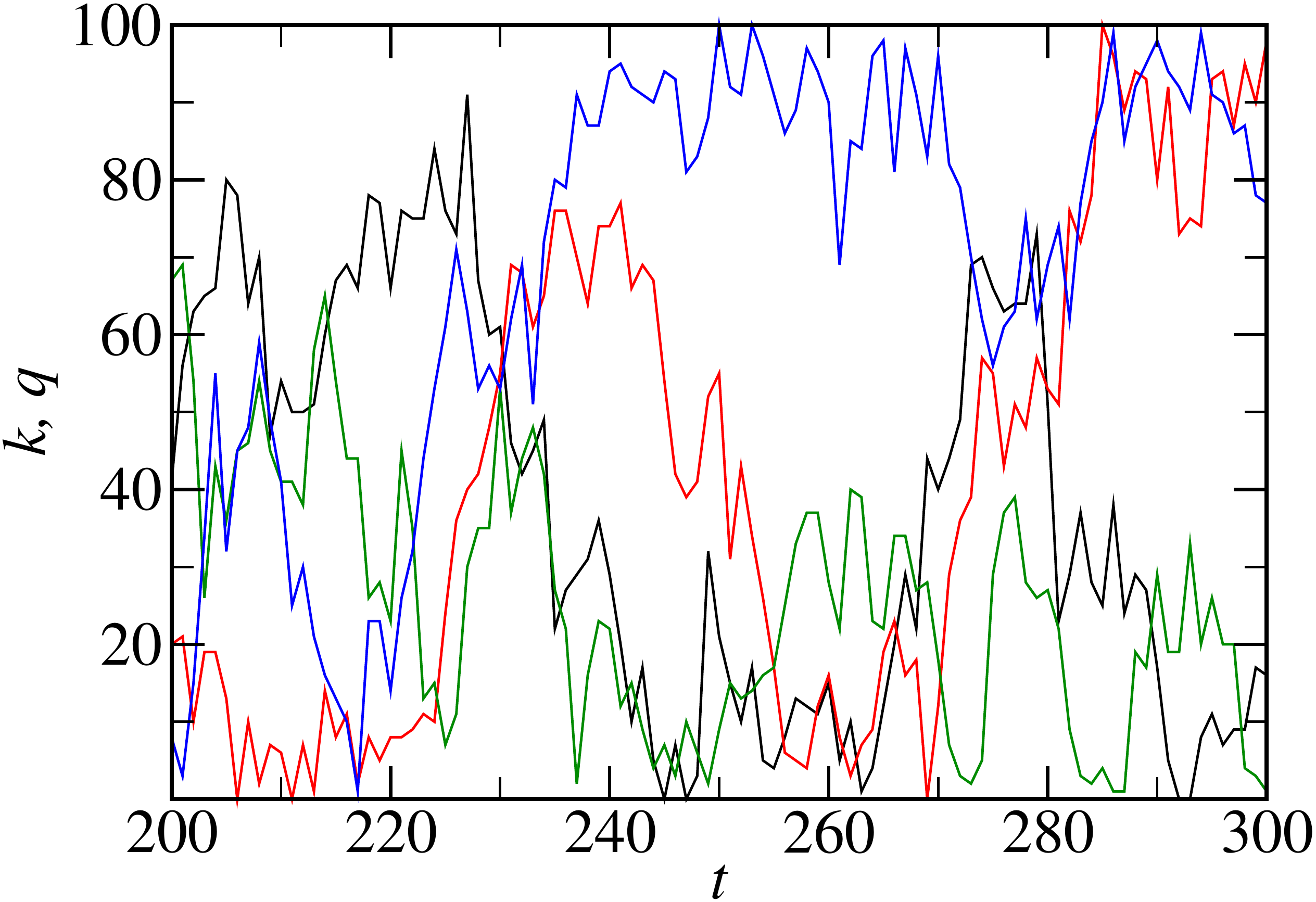}
\caption{An expanded region of Fig. \protect\ref{4minagents}. The four
traces are easily resolved here, showing very different trajectories.}
\label{4minBIG}
\end{figure}

\section{Summary and Outlook}

In the first part of this article, we provided a brief review of networks
with preferred degrees, designed to model a natural social behavior: An
individual tends to reduce or increase its number of contacts if it finds
there are too many or too few. In our baseline model, we assign a fixed
preferred number $\kappa $ to each agent and it cuts/adds a random link if
its degree $k$ is more/less than $\kappa $. The system evolves
stochastically as a random agent is chosen to to act in each attempt.
Despite the apparent randomness of such a network, the degree distributions
display non-trivial properties. Partly due to the detailed balance violating
aspect of its dynamics, the stationary state distribution ($\mathcal{P}
^{\ast }\left( \mathbb{A}\right) $, of the set of networks or adjacency
matrices, $\mathbb{A}$) is not known. Though it is very challenging to
predict averages of observables, some approximation schemes are able to
capture the essence of some quantities, e.g., the degree distribution. Even
more surprising behavior emerges when we study a population with just two
communities ($GIE$ models), differing by the numbers of individuals and
their $\kappa $'s. These puzzles led us to consider an extreme limit, the $%
XIE$ model, with $\kappa =0$ and $\infty $ . Remarkably, detailed balance is
restored here, leading to an exact expression for $\mathcal{P}^{\ast }\left( 
\mathbb{A}\right) $. The system evolves towards `thermal equilibrium' with
an effective Hamiltonian $\mathcal{H}\mathbb{\equiv }-\ln \mathcal{P}^{\ast
} $, though there is no obvious control parameter corresponding to
temperature. In addition, we identified an underlying symmetry, similar to
that in the Ising model. Despite such progress, it is not feasible to
compute, without judicious approximations, the averages of typical
interesting quantities analytically. Through simulations with $N$'s up to $%
3200$, we discover an unusual transition, when $N_{E}/N_{I}$ crosses unity.
Specifically, the network is essentially empty (full) when $N_{E}/N_{I}<1$ ($%
>1$). At the critical point, the value of $X$ wanders over almost the entire
range $\left[ 0,\mathcal{N}\right] $, so that a broad flat plateau appears
in the time averaged distribution, $P(X)$. 
Furthermore, unlike in typical first order transitions, there is no
hysteresis or metastability when $N_{E}-N_{I}$ suddenly change signs. This
astonishing phenomenon, considered an extreme version \cite%
{BarMukamel14,BarMukamel14a} of the Thouless effect, is far from common in
typical statistical systems, the behavior of which generally fall into the
Ehrenfest classification scheme of (first, second, ... order) phase
transitions. 
Focusing on the degree distribution of a single agent, we advance mean field
arguments for their transition probabilities and obtain analytic
predictions. Away from the transition, they are in excellent agreement with
simulation data. At criticality, giant fluctuations render such approaches
ineffective, while few of the system's properties are quantitatively
understood.

In the rest of this article, we present new results on the XIE model and its
novel variants. In the large $N$ limit, we find that the agents of the
majority subgroup are effectively independent. Using both variational and
perturbative techniques, we gain valuable insight into why the mean field
approach is so reliable and find a scaling theory of fluctuations near the
transition. Lastly, we introduce two other variants, the $XIE_{egal}$ and $%
XIE_{elit}$ models, in which the agents show preferential
attachment/detachment. In $XIE_{egal}$, instead of choosing a random
introvert to connect, an `egalitarian' extrovert adds a link to the \textit{%
least} connected $I$. Similarly, an $I$ cuts a link to the \textit{most}
connected $E$. At the opposite extreme is $XIE_{elit}$, in which an
extrovert `elitist' agents adds its link to the most connected $I$, etc.,
modelling those who chase after celebrities. Not surprisingly, while the
egalitarian system leads to a more severe form of the extreme Thouless
effect, the transition for the elitists is continuous. Simulations with
relatively small systems show refreshingly novel behavior, especially for
the elitists. Though these dynamics violate detailed balance and no exact $%
\mathcal{P}^{\ast }$'s are available, we are able to exploit previously used
techniques to obtain good predictions far from criticality. A complete and
quantitative understanding remains to be established.

It is clear that our pursuits raise many interesting questions worthy of
future research. We list only a few here. As pointed out in previous studies 
\cite{LiuSchZia14,BasslerLSZ15}, investigating other properties in the $XIE$
system is clearly a goal within grasp, notably, a systematic study of finite
size scaling will also help to shed light on such peculiarities of this
model. Though of purely theoretical interest, we may treat $\mathcal{H}$ as
a genuine Hamiltonian of the Ising type and introduce (inverse) temperature $%
\beta $ and magnetic field $B$ (i.e., $-\ln z$ above) into a Boltzmann factor%
\begin{equation}
\mathcal{P}\propto \exp \left\{ -\beta \left[ \mathcal{H}-BX\right] \right\}
\end{equation}%
%
%
%
%
The advantage of such a system is that we can study off-critical behavior
not only in the odd-variable $B$, but also in a standard, even-variable $%
\beta $. Properties in an extended phase diagram (as $T$-$H$ for the Ising
model) should shed light on the original $XIE$ model. Finite size scaling
studies can follow standard routes, so that we can follow simple ways to
access critical exponents and scaling functions. In parallel,
renormalization group analysis \cite{BarMukamel14a} should be performed and
the results compared. In this language, we can gain considerable insight
into this system by studying fixed points and their neighborhoods. It will
be very interesting to determine 
if there are relevant operators other than $\beta $ and $B$. 
Finally, identifying the irrelevant variables will let us delineate
clearly the universality class of this $\mathcal{H}$. 

Beyond the $XIE$, its variants and other systems with two communities, we
should explore more realistic populations in which a wide distribution of $%
\kappa $'s is present. Will some of the peculiarities presented above
vanish? or will we face further surprises? Other variants concern a tendency
to favor adding connections to, instead of the celebrity of the time,
`friends of friends.' Correlations will surely build up, while the notion of
communities will emerge in a self-organized manner. Another front involves
assigning weights to the links -- to model the natural tendency for us to
have a few close friends along with many acquaintances \cite{Kaski2011}. It
is easy to devise various quantitative measures of closeness, e.g., how many
minutes per day do two individuals converse. Directed links with weights can
also be introduced -- to account for a speaker's interactions with a large
audience, or the frequency A calls Z \textit{vs.} Z calling A. Having such
weights allows us to impose carrying capacities (e.g., 24 hours per day) in
social networks. In this manner, the portrait of one's friendship is not
restricted to the number of connections: Celebrities may boast 10,000
`friends,' but they cannot converse meaningfully with each one daily!
Needless to say, as we build more realism and complexity into the models, we
should be prepared to face mounting challenges.

Finally, let us emphasize that we focused here only on the topology and
dynamics of the network between individuals with fixed $\kappa $s. The next
natural step is to take into account different states ($\sigma $) an agent
may find itself, e.g., health, wealth, opinion, etc. Indeed, we can expect
an agent's $\kappa $ to depend on not only its $\sigma $ but also those of
others, as illustrated by the example with epidemics in the Introduction.
Beyond social networks, there is considerable interest the behavior of all
types of interacting natural and artificial networks, e.g., ocean currents
and marine food-webs \cite{2}, the internet and the power grid, etc. Such
pursuits should be instrumental as we proceed to the ambitious goal of
understanding adaptive, co-evolving, interdependent networks in general.

\bigskip \bigskip

\ack

We thank M. Barma, C. Del Genio, Ly Do, 
F. Greil, Wenjia Liu, D. Mukamel, B. Schmittmann, Z. Toroczkai for
illuminating discussions and technical assistance. This research is
supported in part by the Indian DST via grant DSTSR/S2/JCB-24/2005, by the
US National Science Foundation via through grants DMR-1206839 and
DMR-1244666, and by AFOSR and DARPA through grant FA9550-12-1-0405. Two of
us (DD,RKPZ) thanks the Galileo Galilei Institute for Theoretical Physics
for hospitality and the INFI for partial support during the summer of 2014.

\bigskip \bigskip

\appendix

\section{Tutorial, using an explicitly solvable case}

For sufficiently small systems, it is possible to find the stationary
distribution and currents by brute force. This section serves as a tutorial
for how the system can be analysed at the most basic level.

The smallest population displaying non-trivial behavior consists of four
nodes, with $\kappa =2$ (i.e., nodes with 2 or 3 links will cut and those
with 0 or 1 link will add). To visualize the graphs, place the nodes on the
corners of a square, so they can be connected via its edges and diagonals.
There are 6 links ($A_{i<j}=1$ or $0$) and so, the configuration space
consists of 64 points: $\left\{
A_{12},A_{13},A_{14},A_{23},A_{24},A_{34}\right\} $.

To find $\mathcal{P}^{\ast }$, we exploit both symmetry and inaccessibility.
The latter accounts for the lack of transitions to the null or complete
graphs, so that we have%
\begin{equation}
\mathcal{P}^{\ast }\left( 0,0,...,0\right) =\mathcal{P}^{\ast }\left(
1,1,...,1\right) =0
\end{equation}%
which reduces the number of unknown $\mathcal{P}^{\ast }$'s to 62. There are
two symmetries, one being permutation. The other is `particle-hole'
symmetry, specific to this $N,\kappa $ combination. Its consequence is that
complimentary graphs share the same $\mathcal{P}^{\ast }$. As a result,
there are just 5 independent $\mathcal{P}^{\ast }$'s. Denoted by $p_{\alpha
} $ ($\alpha =1,...,5$) they correspond to e.g., $\mathcal{P}^{\ast }\left(
1,0,...,0\right) $, $\mathcal{P}^{\ast }\left( 1,1,...,0\right) $, $\mathcal{%
P}^{\ast }\left( 1,0,..,0,1\right) $, $\mathcal{P}^{\ast }\left(
1,0,0,1,0,1\right) $, and $\mathcal{P}^{\ast }\left( 1,1,1,0,0,0\right) $,
respectively.

To account for the symmetries, we introduce $g_{i}$ for their degeneracies.
As an example,%
\begin{eqnarray}
p_{3} &=&\mathcal{P}^{\ast }\left( 1,0,0,0,0,1\right) =\mathcal{P}^{\ast
}\left( 0,1,0,0,1,0\right) =\mathcal{P}^{\ast }\left( 0,0,1,1,0,0\right) \\
&=&\mathcal{P}^{\ast }\left( 0,1,1,1,1,0\right) =\mathcal{P}^{\ast }\left(
1,0,1,1,0,1\right) =\mathcal{P}^{\ast }\left( 1,1,0,0,1,1\right) .
\end{eqnarray}%
implies that $g_{3}=6$. To be clear, we write%
\begin{equation}
g_{\alpha }=6+6,12+12,3+3,4+4,12
\end{equation}%
where the $+$ notation reminds us of the contribution from complimentary
graphs. In the last group , the complimentary graph of any member is also in
the group. The last entry (12) contains graphs Note that these add up to 62.
They serve in the normalization condition%
\begin{equation}
1=\sum_{\alpha }g_{\alpha }p_{\alpha }  \label{norm}
\end{equation}%
and to check probability conservation in a `reduced master equation' for the 
$p$'s. Though symmetries help in reducing the number of unknowns, such a set
of 5 equations must be derived from considering (\ref{MEq}) for specific
configurations. For example, we have%
\begin{eqnarray}
\mathcal{P}^{\ast }\left( 1,0,...,0\right) &=&\frac{1}{6}\mathcal{P}^{\ast
}\left( 0,0,...,0\right) + \\
&&+\frac{1}{8}\left[ 
\begin{array}{c}
\mathcal{P}^{\ast }\left( 1,1,0,0,0,0\right) +\mathcal{P}^{\ast }\left(
1,0,1,0,0,0\right) + \\ 
+\mathcal{P}^{\ast }\left( 1,0,0,1,0,0\right) +\mathcal{P}^{\ast }\left(
1,0,0,0,1,0\right)%
\end{array}%
\right]
\end{eqnarray}%
leading to $p_{1}=\frac{1}{2}p_{2}$. Similar equations for the other 4 $p$'s
can be derived, and we arrive at the `reduced master equation'

\begin{equation}
\left( 
\begin{array}{c}
p_{1} \\ 
p_{2} \\ 
p_{3} \\ 
p_{4} \\ 
p_{5}%
\end{array}%
\right) =\left( 
\begin{array}{ccccc}
0 & 1/2 & 0 & 0 & 0 \\ 
5/12 & 0 & 0 & 1/3 & 1/4 \\ 
1/3 & 0 & 0 & 0 & 1 \\ 
0 & 1 & 0 & 0 & 0 \\ 
0 & 5/6 & 1/2 & 0 & 0%
\end{array}%
\right) \left( 
\begin{array}{c}
p_{1} \\ 
p_{2} \\ 
p_{3} \\ 
p_{4} \\ 
p_{5}%
\end{array}%
\right)  \label{RedME}
\end{equation}%
Since the normalization condition (\ref{norm}) is%
\begin{equation}
1=12p_{1}+24p_{2}+6p_{3}+8p_{4}+12p_{5}
\end{equation}%
the check for probability conservation can be posed as: Is $\left(
12,24,6,8,12\right) $ a left eigenvector with unit eigenvalue for the above
matrix? The answer is indeed \textquotedblleft Yes.\textquotedblright\
Meanwhile, the associated right eigenvector provides us with $%
p_{2}=p_{4}=2p_{1},p_{3}=4p_{1},$ and $p_{5}=11p_{1}/3$. Imposing (\ref{norm}%
), we find%
\begin{equation}
p_{1}=1/144
\end{equation}%
which completes the full stationary distribution.

From this explicit solution, we can compute, e.g., the (average) degree
distribution, $\rho \left( k\right) $. Exploiting symmetry, we can focus on
node 1, say, and study only $\rho \left( 0\right) $ and $\rho \left(
1\right) $. Formally, we write%
\begin{equation}
\rho \left( k\right) \equiv \sum_{\left\{ \mathbb{A}\right\} }\delta \left(
A_{12}+A_{13}+A_{14}-k\right) \mathcal{P}^{\ast }(\mathbb{A})
\end{equation}%
So, 
\begin{equation}
\rho \left( 0\right) =3p_{1}+3p_{2}+p_{4}=\frac{11}{144}=\rho \left( 3\right)
\end{equation}%
A shortcut, namely, $\Sigma _{k}\rho \left( k\right) =1$, can be used to
obtain $\rho \left( 1\right) =$ $61/144=\rho \left( 2\right) $. To check
this result, we verify that $\rho \left( 1\right) =3p_{1}+\left( 6+3\right)
p_{2}+3p_{3}+3p_{4}+6p_{5}$ is indeed $61/144$.

Finally, it is easy to check that detailed balance is violated. For example,
the rate-product is clearly positive for the `elementary loop' involving
graphs of types $1\rightarrow 3\rightarrow 5\rightarrow 2\rightarrow 1$.
Yet, the product for the reversed loop vanishes, as the rate for $%
3\rightarrow 1$ is zero. The Kolmogorov criterion is not satisfied, so that
the system will settle into a \textit{non-equilibrium} stationary state,
with non-trivial stationary probability currents and loops \cite{ZS2007}
(much like those in a steady state electric circuit). The net current, $%
\mathcal{K}^{\ast }$, from $\mathbb{A}$ to $\mathbb{A}^{\prime }$, can be
seen from Eq. (\ref{MEq})%
\begin{equation}
\mathcal{K}^{\ast }(\mathbb{A\rightarrow A}^{\prime })=R(\mathbb{%
A\rightarrow A}^{\prime })\mathcal{P}^{\ast }(\mathbb{A})-R(\mathbb{%
A^{\prime }\rightarrow A})\mathcal{P}^{\ast }(\mathbb{A^{\prime }})
\end{equation}%
The simplest example, since $3\rightarrow 1$ is zero, is $\mathcal{K}^{\ast
}=p_{1}/6=1/864$, for $\mathbb{A}=\left( 1,0,..,0\right) $ and $\mathbb{A}%
^{\prime }=\left( 1,0,..,0,1\right) $. It is instructive to study loops as
well. From this $\mathbb{A}^{\prime }$, we can only transition to one of 4
graphs of the form $\mathbb{A}^{\prime \prime }=\left( 1,1,0,0,0,1\right) $,
and then returning to $\mathbb{A}$\ via $\mathbb{A}^{\prime \prime \prime
}=\left( 1,1,0,0,0,0\right) $. These three $\mathcal{K}$'s are,
respectively, $p_{3}/4-p_{5}/4=1/1728$, $p_{5}/8-5p_{2}/24=1/3456$, and $%
p_{2}/8-5p_{1}/24=1/3456$. As in circuit analysis, there is an instructive
alternative, using loop currents, $\mathcal{I}$, instead. For example, $%
\mathcal{I}^{\ast }(\mathbb{A\rightarrow A}^{\prime }\mathbb{\rightarrow A}%
^{\prime \prime }\mathbb{\rightarrow \mathbb{A}^{\prime \prime \prime
}\rightarrow \mathbb{A}^{\prime }})$ is just $\mathcal{K}^{\ast }(\mathbb{%
A\rightarrow A}^{\prime })/4=1/3456$, since there are four such loops
associated with the $\mathbb{A\rightarrow A}^{\prime }$ segment. A good
exercise is to draw the entire network of configurations and determine all
the loop currents.

\section{Simple, alternative method for determining $\tilde{\protect\rho}$}

Conceptually, it is straightforward that a self consistency condition must
be imposed in the MFA for $\tilde{\rho}_{I,E}$: The input parameters of each
are inextricably linked with the output values of the other. There is a
simpler approach, by considering the probability for $X\rightarrow X\pm 1$
This condition can also be derived from since the averages of each must
satisfy%
\begin{equation}
N_{I}\left\langle k\right\rangle =\left\langle X\right\rangle
=N_{E}\left\langle q\right\rangle
\end{equation}%
Now, $\left\langle k\right\rangle $ can be found from%
\begin{eqnarray}
N_{E}-\left\langle k\right\rangle &=&\sum_{k=0}^{N_{E}}\left( N_{E}-k\right) 
\tilde{\rho}_{I}\left( k\right) =\sum_{\ell =0}^{N_{E}}\frac{\ell \left(
\left\langle p\right\rangle ^{\prime }\right) ^{\ell }}{Z_{I}\ell !} \\
&=&\left\langle p\right\rangle ^{\prime }\left[ 1-\tilde{\rho}_{I}\left(
0\right) \right]
\end{eqnarray}%
But, 
\begin{equation}
\left\langle p\right\rangle ^{\prime }\equiv \left. \sum_{p=1}^{N_{I}}p%
\tilde{\zeta}_{E}\left( p\right) \right/ \sum_{p=1}^{N_{I}}\tilde{\zeta}%
_{E}\left( p\right) =\frac{\left\langle p\right\rangle }{1-\tilde{\zeta}%
_{E}\left( 0\right) }
\end{equation}%
So,%
\begin{eqnarray}
N_{I}\left\langle k\right\rangle &=&N_{I}\left( N_{E}-\left\langle
p\right\rangle \left[ \frac{1-\tilde{\rho}_{I}\left( 0\right) }{1-\tilde{%
\zeta}_{E}\left( 0\right) }\right] \right) \\
&=&N_{E}\left\langle q\right\rangle =N_{E}\left( N_{I}-\left\langle
p\right\rangle \right)
\end{eqnarray}%
giving us%
\begin{equation}
N_{I}\left[ 1-\tilde{\rho}_{I}\left( 0\right) \right] =N_{E}\left[ 1-\tilde{%
\zeta}_{E}\left( 0\right) \right]
\end{equation}%
Note the PHS is manifest here, as well as being automatically satisfied for
the critical case. However, for say, $N_{I}>N_{E}$, we expect $\tilde{\zeta}%
_{E}\left( 0\right) $ to be extremely small and so, arrive at%
\begin{equation}
\tilde{\rho}_{I}\left( 0\right) \cong \frac{\Delta }{N_{I}}  \label{rho0}
\end{equation}%
namely, Eq. (\ref{eq:egal}). Exploiting this and the normalization
conditions, we can find $\left\langle p\right\rangle ^{\prime }$ via%
\begin{equation}
\tilde{\rho}_{I}\left( 0\right) \left\{ 1+\frac{N_{E}}{\left\langle
p\right\rangle ^{\prime }}+\frac{N_{E}}{\left\langle p\right\rangle ^{\prime
}}\frac{N_{E}-1}{\left\langle p\right\rangle ^{\prime }}+...\right\} =1
\label{p'eq}
\end{equation}%
This equation can be written in closed form with the help of the exponential
sum function, $e_{n}\left( \xi \right) \equiv \Sigma _{0}^{n}\xi ^{\ell
}/\ell !$, or an incomplete $\Gamma $ function, but it is simpler to
determine $\left\langle p\right\rangle ^{\prime }$ numerically using (\ref%
{p'eq}). For specific cases, we have checked that this approach indeed
produces $\tilde{\zeta}_{E}\left( 0\right) \ll 1$, which justifies the use
of (\ref{rho0}).

\section{Truncated Poisson distribution}

In this appendix, we provide details of our distributions (\ref{tilde-dists}%
). In the literature (e.g., \cite{Moore1954}), this kind of truncation is
known as Type 1. We also dealt with $\left\langle \bullet \right\rangle
^{\prime }$, which comes under the heading of Type 3 truncated Poisson
distributions. Defined on $\ell \in \left[ 0,n\right] $ 
\begin{equation}
T\left( \ell ;\xi ,n\right) =\frac{\xi ^{\ell }}{e_{n}\left( \xi \right)
\ell !}
\end{equation}%
is proportional to the standard Poisson distribution within its support. In
the limit of $n\rightarrow \infty $ with fixed $\xi $, $T$ reduces trivially
to the Poisson: $e^{-\xi }\xi ^{\ell }/\ell !$. However, if both $\xi $ and $%
n\rightarrow \infty $ , with fixed $\xi /n\equiv \mu >1$, then it is clear
that $T$ increases monotonically in $\left[ 0,n\right] $. Since $T$ peaks at 
$\ell =n$, it is natural to use the variable $\bar{\ell}\equiv n-\ell $ and
to study%
\begin{equation}
T\left( \ell ;\xi ,n\right) =\frac{\xi ^{n-\bar{\ell}}}{e_{n}\left( \xi
\right) \left( n-\bar{\ell}\right) !}  \label{T}
\end{equation}%
in the regime of $\bar{\ell}\thicksim O\left( 1\right) $. Exploiting (\ref{F}%
-\ref{Fapprox}), we arrive at the exponential%
\begin{equation}
T\left( \ell ;\xi ,n\right) \propto \mu ^{-\bar{\ell}}~~.  \label{Texp}
\end{equation}

\newpage


\end{document}